\documentclass[twocolumns,structabstract]{aa}
\usepackage[latin1]{inputenc}
\usepackage[dvips]{graphicx}
\usepackage{amssymb}
\usepackage{amsmath}
\usepackage{txfonts}
\usepackage[comma,authoryear]{natbib}
\usepackage{hyperref}

\newcommand{\ie}{\textit{i.e.}}
\newcommand{\eg}{\textit{e.g.}}

\newcommand{\OmegaK}{\Omega_\mathrm{K}}
\newcommand{\OmegaC}{\Omega_\mathrm{C}}

\newcommand{\dpart}[2]{\frac{\partial #1}{\partial #2}}

\newcommand{\vect}{\vec}

\renewcommand{\l}{\ell}

\newcommand{\eo}{\vect{e}_{\mathrm{obs.}}}

\newcommand{\geff}{g_{\mathrm{eff}}}
\newcommand{\Teff}{T_{\mathrm{eff}}}

\begin{document}

\title{Frequency regularities of acoustic modes and multi-colour mode
identification in rapidly rotating stars}
\author{D. R. Reese\inst{1,2,3}
        \and
        F. Lignières\inst{4,5}
        \and
        J. Ballot\inst{4,5}
        \and
        M.-A. Dupret\inst{1}
        \and
        C. Barban\inst{3}
        \and
        C. van~'t Veer-Menneret\inst{6}
        \and
        K. B. MacGregor\inst{7}
       }

\institute{Institut d'Astrophysique et Géophysique de l'Université de Liège,
           Allée du 6 Août 17, 4000 Liège, Belgium
           \and
           School of Physics and Astronomy, University of Birmingham, Edgbaston, Birmingham, B15 2TT, UK
           \and
           LESIA, Observatoire de Paris, PSL Research University, CNRS, Sorbonne
           Universités, UPMC Univ. Paris 06, Univ. Paris Diderot, Sorbonne Paris
           Cité, 5 place Jules Janssen, 92195 Meudon, France
           \email{daniel.reese@obspm.fr}
           \and
           Universit{\'e} de Toulouse, UPS-OMP, IRAP, Toulouse, France
           \and
           CNRS, IRAP, 14 avenue Edouard Belin, 31400 Toulouse, France
           \and
           GEPI, Observatoire de Paris-Meudon, CNRS, Université Paris Diderot, 92125, Meudon Cedex, France
           \and
           High Altitude Observatory, National Center for Atmospheric Research, Boulder, CO 80307, USA
          }
\date{}

\abstract
{Mode identification has remained a major obstacle in the interpretation of
pulsation spectra in rapidly rotating stars.  This has motivated recent work on
calculating realistic multi-colour mode visibilities in such stars.}
{We would like to test mode identification methods and seismic diagnostics in
rapidly rotating stars, using oscillation spectra based on these new theoretical
predictions.}
{We investigate the auto-correlation function and Fourier transform of
theoretically calculated frequency spectra, in which modes are selected
according to their visibilities.  Given that intrinsic mode amplitudes
are determined by non-linear saturation and cannot currently be theoretically
predicted, we experimented with various ad-hoc prescriptions for setting the
mode amplitudes, including using random values. Furthermore, we analyse the
ratios between mode amplitudes observed in different photometric bands to see up
to what extent they can identify modes.}
{When non-random intrinsic mode amplitudes are used, our results show
that it is possible to extract a mean value for the large frequency separation
or half its value, and sometimes twice the rotation rate, from the
auto-correlation of the frequency spectra.  Furthermore, the Fourier transforms
are mostly sensitive to the large frequency separation or half its value. The
combination of the two methods may therefore measure and distinguish the two
types of separations.  When the intrinsic mode amplitudes include random
factors, as seems more representative of real stars, the results are far less
favourable.  It is only when the large separation or half its value coincides
with twice the rotation rate, that it might be possible to detect the signature
of a frequency regularity.  We also find that amplitude ratios is a good way of
grouping together modes with similar characteristics.  By analysing the
frequencies of these groups, it is possible to constrain mode identification as
well as determine the large frequency separation and the rotation rate.}
{}


\maketitle

\section{Introduction}

One of the major obstacles in interpreting the acoustic frequency
spectra of rapidly rotating stars is mode identification, \ie\ finding the
correspondence between theoretically calculated modes and observed pulsations. 
Several reasons make it difficult to match the two.  First and foremost is the
the lack of simple frequency patterns, such as what is present in solar-like
stars.  Indeed, rapid rotation leads to complex spectra with overlapping classes
of pulsation modes, each with an independent frequency organisation
\citep{Lignieres2008}.  Next comes the whole problem of mode amplitudes.  Most
rapid rotators tend to be massive or intermediate mass stars where modes are
predominantly excited by the $\kappa$ mechanism.  This leads to non-linear
saturation and coupling between modes, making it nearly impossible to
predict the amplitudes with current theory.  Further difficulties include
avoided crossings between modes, and all of the theoretical and numerical
challenges associated with rapid rotation.  From an observational point of view,
the high quality data from space missions CoRoT \citep{Baglin2009, Auvergne2009}
and Kepler \citep{Borucki2009} have painted a new picture of $\delta$ Scuti
stars, through the detection of hundreds of pulsation modes \citep{Poretti2009,
Balona2012}.  In a similar way, the number of detected modes has also increased
for stars from other classes of rapidly rotating pulsators, and along with it
the complexity of the spectra \citep[\eg][]{Uytterhoeven2011}.  Consequently,
most asteroseismic analyses have focused on interpreting the general
characteristics of these spectra rather than identifying individual modes.

Various strategies have been devised in order to identify modes.  One can, for
instance, search for frequency patterns appropriate for rapid rotation. The
background for this search is the discovery of asymptotically uniform frequency
spacings  in the numerically computed spectra of uniformly rotating polytropic
models \citep{Lignieres2006, Reese2008a} and differentially rotating realistic
SCF models \citep{Reese2009a}. These uniform spacings have also been modelled
through asymptotic semi-analytical formulas \citep{Pasek2012}. In observed
spectra, recurrent frequency spacings which may correspond to the large
separation or half its value have been found in some stars
\citep{GarciaHernandez2009, GarciaHernandez2013, Paparo2016}.  Moreover,
\citet{GarciaHernandez2015} showed that mean density estimates based on such
spacings \citep[obtained via a scaling relation similar to the one in][but based
on SCF models]{Reese2008a} are compatible with independent mass and radii
measurements obtained for $\delta$ Scuti stars in binary systems.  Nonetheless,
it is expected that various effects should contribute to hide these regular
frequency patterns. First, as mentioned before, the full spectrum is a
superposition of sub-spectra corresponding to different classes of modes and
some of the uniform spacings only concern one class.  This complicates their
detection in the full spectrum.  Also, due to their asymptotic nature, these
spacings might not be relevant to analyse the low to moderate (up to radial
order $n \sim 10$) frequency domain, typical of most rapidly rotating pulsators.
A third effect that might come into play is the presence of mixed modes in
evolved stars and/or sharp sound speed gradients as they can potentially modify
the regular spacings.  Finally, mode selection effects due to the non-linearly
determined intrinsic mode amplitudes could affect the detectability of the
regular patterns. As a first attempt, \citet{Reese2009b} developed a strategy to
find these frequency spacings but ran into difficulties when including chaotic
modes, which come from another class of modes. \citet{Lignieres2010} addressed
the same question with encouraging results but their analysis was restricted to
the asymptotic regime and relied on simplifying assumptions regarding the
spectrum of chaotic modes and the mode visibilities. In the present paper, our
first goal is to search for regular frequency spacings in the most realistic
synthetic spectra available, using relevant frequency ranges and accurate
visibility calculations.  While they can provide guidance to a similar search in
real data, we already know that these results must be taken with
caution since the intrinsic mode amplitudes used in the present paper
are not realistic but based on ad-hoc prescriptions.

Another strategy which avoids this difficulty is to constrain the identification
through multi-colour photometric or spectroscopic observations.  Indeed, one can
measure the amplitudes and the phases of a given pulsation mode in different
photometric bands and then compare these by calculating amplitude ratios or
phase differences.  The geometry of the modes will then lead to different
characteristic signatures.  \textit{One important advantage of this method is
that these signatures are independent of the intrinsic mode amplitude.}  In a
similar fashion, the oscillatory movements induced by a pulsation mode cause
Doppler shifts which show up as variations in the shape of spectroscopic
absorption lines, known as ``line profile variations'' or LPVs.  These
variations are then directly related to the geometry of the mode.  By comparing
theoretical predictions with observations, one can then constrain the mode's
identification.  In what follows, we will focus on multi-colour mode
identification.

Most of the previous theoretical investigations of amplitude ratios and phase
differences have been based on mode calculations which approximate the effects
of rotation.  For instance, \citet{Daszynska_Daszkiewicz2002} used a
perturbative approach, whereas \citet{Townsend2003b} and
\citet{Daszynska_Daszkiewicz2007} applied the traditional approximation
(which is typically used when calculating gravito-inertial modes).  It
is only recently that such predictions have started to fully take into account
the effects of rotation.  First, \citet{Lignieres2006} and \citet{Lignieres2009}
calculated disk-integration factors for pulsation modes calculated in fully
deformed polytropic models.  Given the simplified nature of these calculations,
it was not possible to calculate associated amplitude ratios or phase
differences. More recently, \citet[][hereafter Paper I]{Reese2013} calculated
mode visibilities in realistic models of rapidly rotating stars. This work
relied on a grid of Kurucz atmospheres to calculate realistic emerging
intensities, thereby taking into account limb and gravity darkening.  It took
into account the Lagrangian variations of temperature and effective gravity as
well as the surface distortion from the pulsation modes. The pulsation modes
were calculated using a 2D approach which takes into account centrifugal
distortion.  The main limitation was the adiabatic approximation which leads to
an unreliable estimation of the Lagrangian variations of the effective
temperature.  Nonetheless, first results were obtained in that work which
hopefully provide a qualitative insight, both into mode visibilities and
multi-colour amplitude ratios.

In the following two sections, we succinctly recall some of the main aspects of
pulsation modes in rapidly rotating stars, as well as the basic principles
behind the visibility calculations described in Paper~I.  We then examine the
auto-correlation functions of theoretically calculated spectra which overlap the
p- and g-mode domains.  This is followed by a discussion on Fourier transforms
of frequency spectra and how these complement auto-correlation functions.  In
Section~\ref{sect:multicolour}, we show how multi-colour photometric mode
identification can be extended to rapidly rotating stars.  Finally, a discussion
concludes the paper.

\section{Pulsation modes}
\label{sect:modes}

Rapidly rotating models are calculated thanks to the Self-Consistent Field (SCF)
method \citep{Jackson2005, MacGregor2007}.  The resultant models represent ZAMS
stars with a cylindrical rotation profile, and hence a barotropic stellar
structure, \ie\ a structure where different thermodynamic quantities such as the
density, pressure, and temperature are constant on isopotential surfaces
(deduced from the sum of the gravitational and centrifugal potentials).  The
pulsation modes are calculated using the Two-dimensional Oscillation Program
\citep[TOP][]{Reese2006,Reese2009a}.  As described in Paper I, an improved
treatment of the models and a slightly different mechanical boundary condition
were necessary to obtain eigenfunctions appropriate for visibility calculations.
Compared to Paper~I, the frequency range has been extended both to higher and
lower frequencies, including g-modes, although for the most part we focus on
p-modes.

\begin{figure}
\begin{center}
\begin{tabular}{cc}
\textbf{Odd} &
\textbf{Even} \\
 \includegraphics[width=0.23\textwidth]{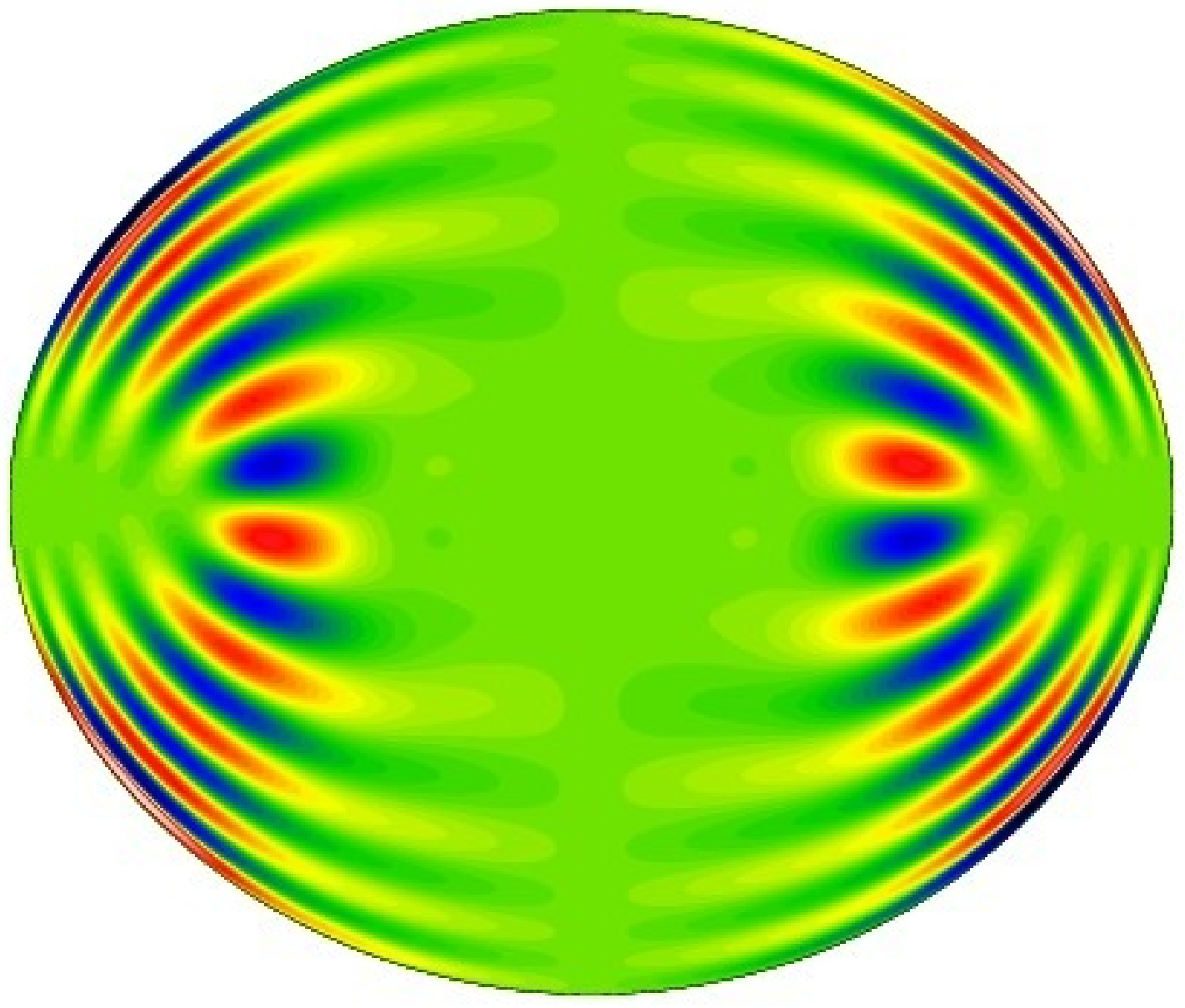} &
 \includegraphics[width=0.23\textwidth]{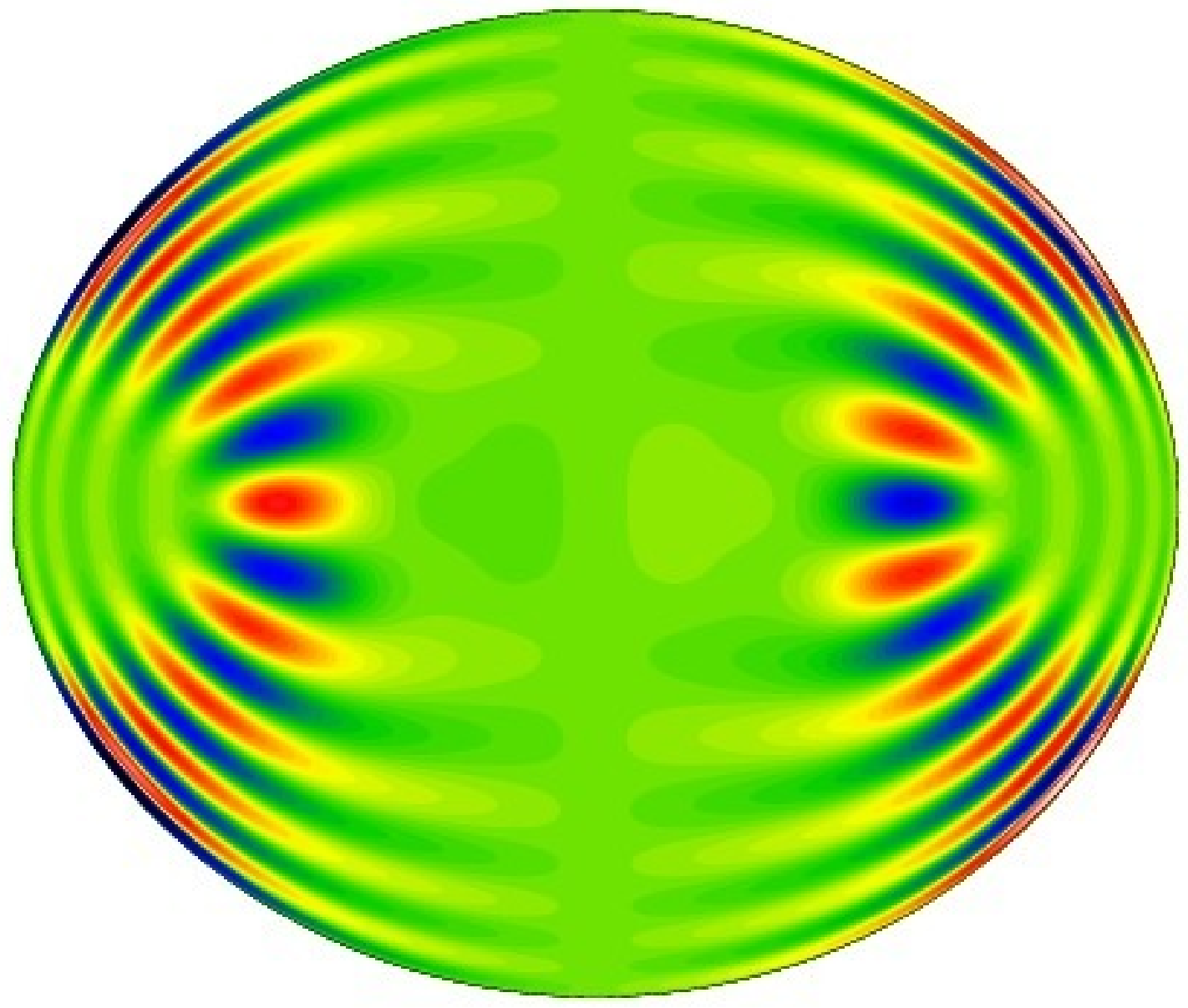} \\
$(n,\,\l,\,m)=(9,2,1)$ &
$(n,\,\l,\,m)=(10,1,1)$ \\
$(\tilde{n},\,\tilde{\l},\,m)=(19,0,1)$ &
$(\tilde{n},\,\tilde{\l},\,m)=(20,0,1)$ 
\end{tabular}
\end{center}
\caption{(colour online) Meridional cross-sections of two island modes with the
same $(\tilde{\l},\,m)$ values but opposing parity, as determined from the
parity of $\tilde{n}$.  The colours indicate the Eulerian pressure perturbation
divided by the square root of the equilibrium density. \label{fig:parity}}
\end{figure}

As was shown in \citet{Lignieres2008, Lignieres2009}, acoustic modes subdivide
into several classes of modes as the rotation rate increases.  Each class of
mode has its own typical geometry and frequency organisation, whether regular or
statistical.  This behaviour stems from the gradual transition of the ray
dynamics system from being integrable to chaotic -- a transition which causes
different regions, associated with the different classes, to appear in the
Poincaré section.  Of particular interest are the island modes, the rotating
counterpart to low degree modes.  These modes focus around a periodic trajectory
and are characterised by the quantum numbers $(\tilde{n},\,\tilde{\l},\,m)$
where $\tilde{n}$ is the number of nodes along the trajectory, $\tilde{\l}$ the
number of nodes perpendicular to the trajectory, and $m$ the usual azimuthal
order.  As illustrated in the animation in \citet{Reese2008b}, it is possible to
transform these quantum numbers into the usual spherical quantum numbers
$(n,\,\l,\,m)$ and inversely, using the following relations based on what could
be described as ``node conservation'':
\begin{equation}
\left\{
\begin{array}{rcl}
\tilde{n}   &=&      2n + \varepsilon, \\
\tilde{\l}  &=&     \frac{\l - |m| - \varepsilon}{2}, \\
\varepsilon &\equiv& (\l+m)\,\, [2],
\end{array}
\right. \qquad
\left\{
\begin{array}{rcl}
n &=& \frac{\tilde{n} - \varepsilon}{2}, \\
\l &=& 2\tilde{\l} + |m| + \varepsilon, \\
\varepsilon &\equiv& \tilde{n}\,\, [2],
\end{array}
\right.
\end{equation}
where $\varepsilon$ corresponds to mode parity, \ie\ $\varepsilon=0$ for even
modes, modes which are symmetric with respect to the equator, and
$\varepsilon=1$ for odd modes.  One will note in particular that the parity of
$\tilde{n}$ corresponds to that of the mode.  This simply corresponds to the
fact that the node along the equatorial plane is treated as a pseudo-radial node
in island modes.  Closely related to this is the fact that one set of
$(\tilde{\l},\,m)$ values corresponds to two sets of $(\l,\,m)$ values depending
on the parity of $\tilde{n}$, as illustrated in Fig.~\ref{fig:parity}.  At this
point, it is also useful to introduce the large frequency separation, $\Delta$
and the semi-large frequency separation, $\Delta/2$:
\begin{eqnarray}
\label{eq:delta}
\Delta &=& \omega_{n+1,\,\l,\,m} - \omega_{n,\,\l,\,m}
        =  \omega_{\tilde{n}+2,\,\tilde{\l},\,m}
        -  \omega_{\tilde{n},\,\tilde{\l},\,m}, \\
\label{eq:delta_2}
\frac{\Delta}{2} &=& \omega_{\tilde{n}+1,\,\tilde{\l},\,m} 
                  -  \omega_{\tilde{n},\,\tilde{\l},\,m},
\end{eqnarray}
where $\omega$ is the frequency, indexed either by the spherical or island mode
quantum numbers.

Recently, \citet{Pasek2012} found asymptotic expressions for the profile of
island modes in the direction  perpendicular to the periodic trajectory.  Apart
from a $1/\sqrt{\omega}$ scaling, modes with the same $(\tilde{\l},\,m)$ values
have the same transverse profile.   This applies in particular at the stellar
surface.  Hence, modes with the same $(\tilde{\l},\,m)$ values will have similar
surface profiles \textit{in one hemisphere}. They will be either symmetric or
antisymmetric with respect to the equator depending on the parity of
$\tilde{n}$. Note that using $\l$ instead of $\tilde{\l}$ also enables
us to select island modes of the same equatorial parity. In addition, the weak
influence of the Coriolis force on high-frequency acoustic modes
\citep[see][]{Reese2006} implies that $\pm m$ pairs of mode are nearly identical
(with near-degenerate frequencies in the corotating frame).  It follows that
island modes with the same $(\l,\,|m|)$ values will have similar surface
profiles. They should then have  similar visibilities (as described in the next
section) and this property will play an important role in the mode
identification methods described in what follows.

\section{Mode visibilities}
\label{sect:visi}
 
Mode visibilities are obtained by perturbing the expression which gives the
amount of energy radiated by a star to an observing instrument:
\begin{equation}
E = \frac{1}{2\pi d^2}\iint_{\mathrm{Vis. Surf.}} I(\mu,\geff,\Teff) \eo \cdot \vect{\mathrm{d}S}
\label{eq:E}
\end{equation}
where $d$ is the distance to the star, $\mu$ the cosine of the angle between the
outward normal to the stellar surface and the observer's direction, ``Vis.
Surf.'' the surface visible to the observer, $\geff$ and $\Teff$ the effective
gravity and temperature, and $I(\mu,\geff,\Teff)$ the specific radiation
intensity, multiplied by the instrument's and/or filter's transmission curve and
integrated over the wavelength spectrum.  The perturbed expression is:
\begin{eqnarray}
\Delta E(t) &=& \frac{1}{2\pi d^2}\Re \left\{
              \iint_{\mathrm{Vis. Surf.}}  \delta I(\mu,\geff,\Teff,t) \eo \cdot \vect{\mathrm{d}S} \right.  \nonumber \\
            & & \left. + \iint_{\mathrm{Vis. Surf.}}  I(\mu,\geff,\Teff)  \eo \cdot \delta(\vect{\mathrm{d}S}) \right\}
\label{eq:E_perturbed}
\end{eqnarray}
where $\delta$ denotes a Lagrangian perturbation and $\Re\{ \dots \}$ the real
part.  Variations caused by fluctuations to the boundary between the visible and
hidden side of the star lead to second order effects and are therefore
neglected.  The Lagrangian perturbation to the specific intensity, $\delta I$,
is calculated as follows:
\begin{equation}
\delta I = I \left(\dpart{\ln I}{\ln \Teff} \frac{\delta \Teff}{\Teff}
         + \dpart{\ln I}{\ln \geff} \frac{\delta \geff}{\geff} \right)
         + \dpart{I}{\mu} \delta \mu
\end{equation}
The quantities $I$, $\partial \ln I/ \partial \ln \Teff$ , $\partial \ln
I/\partial \ln \geff$ and $\partial I/\partial \mu$ are calculated using a grid
of Kurucz atmospheres which span the relevant effective temperature and gravity
ranges (see references and more details in Paper I).  This allows us to take
into account both limb and gravity darkening.  The quantities $\delta
\Teff/\Teff$, $\delta \geff/\geff$ and $\delta \mu$ are deduced from the surface
profiles of the pulsation modes as described in Paper~I.  We note that since the
pulsation modes are calculated using the adiabatic approximation, $\delta
\Teff/\Teff$ is not accessible and is therefore approximated by $\delta T/T$. As
has been pointed out in \citet{Dupret2002} and \citet{Dupret2003}, this can lead
to poor results, since $\delta T/T$ is not reliable in the outer layers when
calculated adiabatically.  A full non-adiabatic calculation would remedy this
problem but is beyond the scope of this paper. Finally, the term
$\delta(\vect{\mathrm{d}S})$, which intervenes in the second integral in
Eq.~(\ref{eq:E_perturbed}) is also deduced from the surface profiles of the
eigenmodes, as described in Paper~I.  Hence, the geometrical distortions of the
stellar surface, induced by the pulsation modes are fully taken into account. 
We note that the centrifugal deformation is also taken into account, both in the
pulsation and visibility calculations.

\section{Auto-correlation function of the frequency spectra}
\label{sect:auto}

\subsection{General description}
\label{sect:normalisations}

We first look at the auto-correlation function of the frequency spectra using
mode visibilities in CoRoT's photometric band to set the amplitudes.  These
spectra have been calculated in 2 $M_{\odot}$ stellar models with rotation rates
ranging from $0.0$ to $0.8\,\OmegaC$, where $\OmegaC$ is the critical rotation
rate\footnote{The critical rotation rate described here differs from that
typically used by observers.  Indeed, it is calculated using
$g_{\mathrm{eq}}$, the equatorial gravity of the current model, whereas
most observers use the equatorial gravity of the model at breakup,
which tends to be smaller due to the increased equatorial radius.  As such, the
values used here convert to larger values if using the observers' convention.},
the rotation rate at which the centrifugal force exactly compensates the
equatorial gravity,  $g_{\mathrm{eq}}$ \citep[\eg][]{Jackson2005}:
\begin{equation}
\OmegaC = \sqrt{\frac{g_{\mathrm{eq}}}{r}}.
\end{equation}

We extract the $N$ most visible modes in the CoRoT photometric band. 
The selected frequencies are given the same amplitude and are convolved
by a Gaussian profile with a width of 1/15 d$^{-1}$, after which we
calculate the auto-correlation function.  The width of the Gaussian profile had
to be carefully selected.  Indeed, a smaller width leads to signatures which
are less clear given the variations of the large frequency separation in
the frequency range considered here, whereas a larger width leads to a loss of
accuracy in the position of peaks in the auto-correlation function.

In order to select modes according to how visible they are, one not only
needs their visibilities (as computed in Sect.~\ref{sect:visi}) but also their
intrinsic amplitudes, since the observed amplitude is proportional to the
product of the two.  However, determining the intrinsic amplitude of a mode in a
classical pulsator is an unsolved theoretical problem \citep[][and references
therein]{Goupil2005}. Accordingly, we will experiment with the following ad-hoc
ways of defining the intrinsic amplitude:
\begin{itemize}
\item normalisation of the maximal displacement:
\begin{equation}
(\omega+m\Omega)^2 \max_{V} \left\|\vect{\xi}\right\| = \mbox{constant}
\label{eq:normalisation_disp}
\end{equation}
\item inclusion of random factors:  the mode visibilities from the
previous case are multiplied by random numbers.
\item normalisation of the kinetic energy:
\begin{equation}
(\omega+m\Omega)^3 \sqrt{\int_V \rho_0 \|\vect{\xi}\|^2 \mathrm{d}V} = \mbox{constant}
\label{eq:normalisation_Ek}
\end{equation}
\item normalisation of the mean surface displacement:
\begin{equation}
(\omega+m\Omega)^2 \sqrt{\int_S \|\vect{\xi}(R)\|^2 \mathrm{d}S} = \mbox{constant}
\label{eq:normalisation_surf}
\end{equation}
\end{itemize}
In each case, an extra power of $(\omega+m\Omega)$ is introduced in
order to have a near constant amplitude for acoustic modes of the same degree
and increasing frequency (as would be the case in a slowly rotating solar-like
pulsator). In the first three cases, the normalisation by a volume-related
quantity ensures that gravity (or gravito-inertial) modes have lower surface
amplitudes than acoustic modes. In contrast, gravity modes  are not penalised
when the mode amplitudes are determined by the mean surface displacement
normalisation. We also experiment a case where the intrinsic mode amplitudes are
multiplied by a random factor. This enables us to test how the auto-correlation
function is affected by a drastic reordering of the observed amplitudes
(which could occur as a result of non-linear interactions between
modes).

\subsection{Normalisation of the maximal displacement}

\begin{figure*}
\begin{center}
 \includegraphics[width=0.8\textwidth]{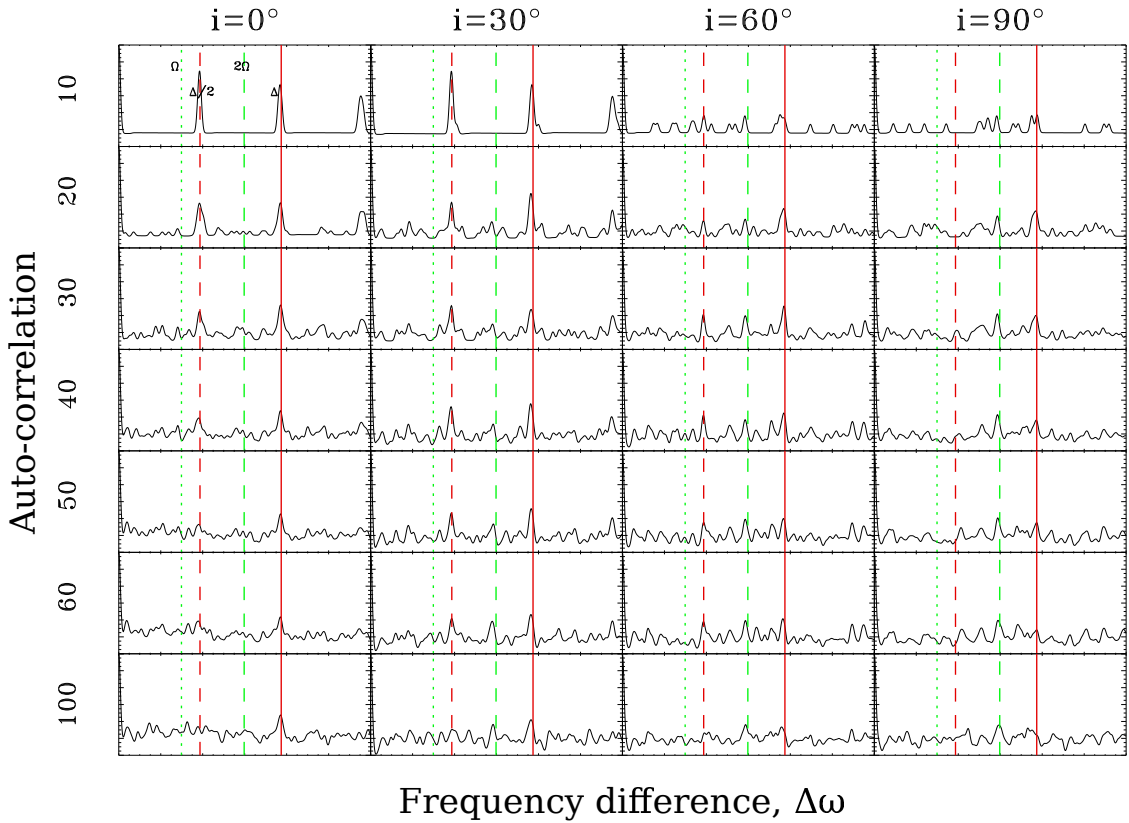}
\end{center}
\caption{(colour online) Auto-correlation functions of acoustic spectra in a 2
$M_{\odot}$ stellar model rotating at $0.5\,\OmegaC$.  The spherical radial
orders of the modes, $n$, range from $2$ to $9$, therefore spanning $7$
large frequency separations.  The four columns correspond to four different
inclinations, $i=0^{\circ}$ being a pole-on configuration.  Each row corresponds
to a different number of included modes.  For example, in the top row, only the
$10$ most visible modes are included in the frequency spectra before calculating
the auto-correlation function.  The vertical dotted and dashed green lines give
the rotation rate, $\Omega$, and twice its value.  The vertical dashed and
continuous red lines indicate the large frequency separation, $\Delta$, and half
its value (see Eqs.~\ref{eq:delta} and~\ref{eq:delta_2}). \label{fig:auto5_7D}}
\end{figure*}

Figure~\ref{fig:auto5_7D} shows the auto-correlation functions of a
spectrum spanning $7$ large frequency separations from $(n,\,\l,\,m)=(2,1,0)$ to
$(9,1,0)$, corresponding to a 2 $M_{\odot}$ stellar model rotating at
$0.5\,\OmegaC$. Auto-correlation functions have been computed for four different
inclination angles and for various amplitude thresholds decreasing from $N=10$
to $100$.  We note that for the pole-on configuration $i=0^{\circ}$, only
axisymmetric modes are visible.  Accordingly, high values of $N$ implicitly lead
to the assumption that $\l > 3$ modes are visible which may be somewhat
optimistic and is subject to caution (although some publications do suggest such
modes can sometimes be detected, see, \eg\ \citealt{Poretti2009}).  Very clear
signatures of the large frequency separation, $\Delta$, and half its value show
up for $i=0^{\circ}$ and $i=30^{\circ}$.  These signatures are caused by the
dominant presence of island modes.  We also observe that the $\Delta/2$
signature disappears at large inclination angles (see the fourth column in
Fig.~\ref{fig:auto5_7D}). This is due to the cancellation of antisymmetric
island modes seen in near equator-on configurations.

Another important feature of the auto-correlation function is a small
peak at twice the rotation rate mostly seen for high values of $i$ and using
large numbers of selected modes. This is caused by the frequency difference
between prograde modes with $m=-1$ and their retrograde counterparts, $m=1$. 
Indeed, as already mentioned, the weak effect of the Coriolis force induces a
near-degeneracy of $\pm m$ modes in the corotating frame, regardless of whether
they are island or chaotic in  nature.  This produces pairs of frequencies
separated by $2m\Omega$ in the observed spectra of a uniformly rotating star.
Their visibilities are also very similar. The $2\Omega$ peak is then due to
$m=\pm1$ pairs, the most visible non-axisymmetric modes.

\begin{figure*}
\begin{center}
 \includegraphics[width=0.8\textwidth]{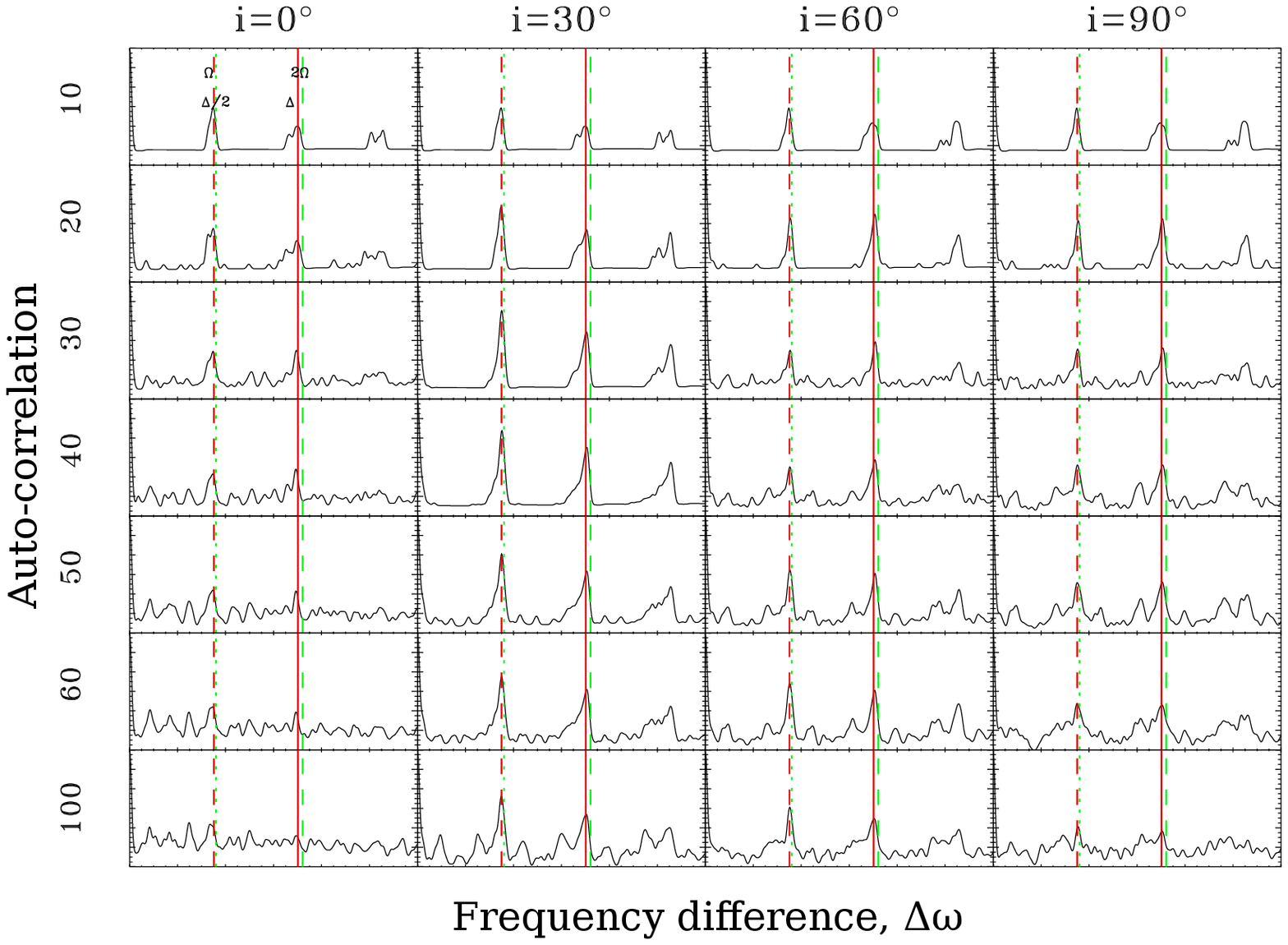}
\end{center}
\caption{(colour online) Same as Fig.~\ref{fig:auto5_7D} except that the model
is rotating at $0.7\,\OmegaC$. \label{fig:auto7_7D}}
\end{figure*}

Figure~\ref{fig:auto7_7D} shows what happens with a model rotating at
$0.7\,\OmegaC$.  This time, the large frequency separation is very
similar to twice the rotation rate.  As a result, the auto-correlation
functions show very strong peaks at both $\Delta \simeq 2\Omega$ and $\Delta/2
\simeq \Omega$, regardless of the inclination angle.  This is not
surprising as the corresponding regularities add up to produce strong peaks. 
The pulsation frequencies actually tend to cluster around points separated  by
$\Delta/2 \simeq \Omega$. While such regularities are then easier to spot, it is
difficult to disentangle between changing the pseudo-radial order, $\tilde{n}$, 
and changing the azimuthal order.  A similar coincidence occurs around
$0.3\OmegaC$ where $2\Omega$ is close to $\Delta/2$.  This also leads to strong
peaks in the auto-correlation functions.

It is interesting to observe that the frequency spacing $\Omega \simeq \Delta/2$
shows up quite strongly in Fig.~\ref{fig:auto7_7D}, even in the equator-on
($i=90^{\circ}$) configuration.  This is somewhat surprising since
antisymmetric modes cancel out, only leaving even modes which are
spaced by $\Delta$ for fixed values of $(\tilde{\l},\,m)$. Furthermore, prograde
and retrograde modes with $m=\pm 1$ are spaced by $2\Omega$.  The explanation
lies in the fact that if one considers a ``multiplet'' of modes with the same
$(\tilde{n},\,\tilde{\l})$ values, at sufficient rotation rates, modes with
consecutive $m$ values are approximately separated by $\Omega$, at least for
small values of $|m|$.  Indeed, the advection term $m\Omega$ is much stronger
than the frequency deviations in the corotating frame which behaves as 
$m/\sqrt{\tilde{n}}$ according to the numerical calculations of
\citet{Reese2009a} and the analytical model of \citet{Pasek2012}.

\subsection{Inclusion of random factors}

\begin{figure*}
\begin{center}
 \includegraphics[width=0.8\textwidth]{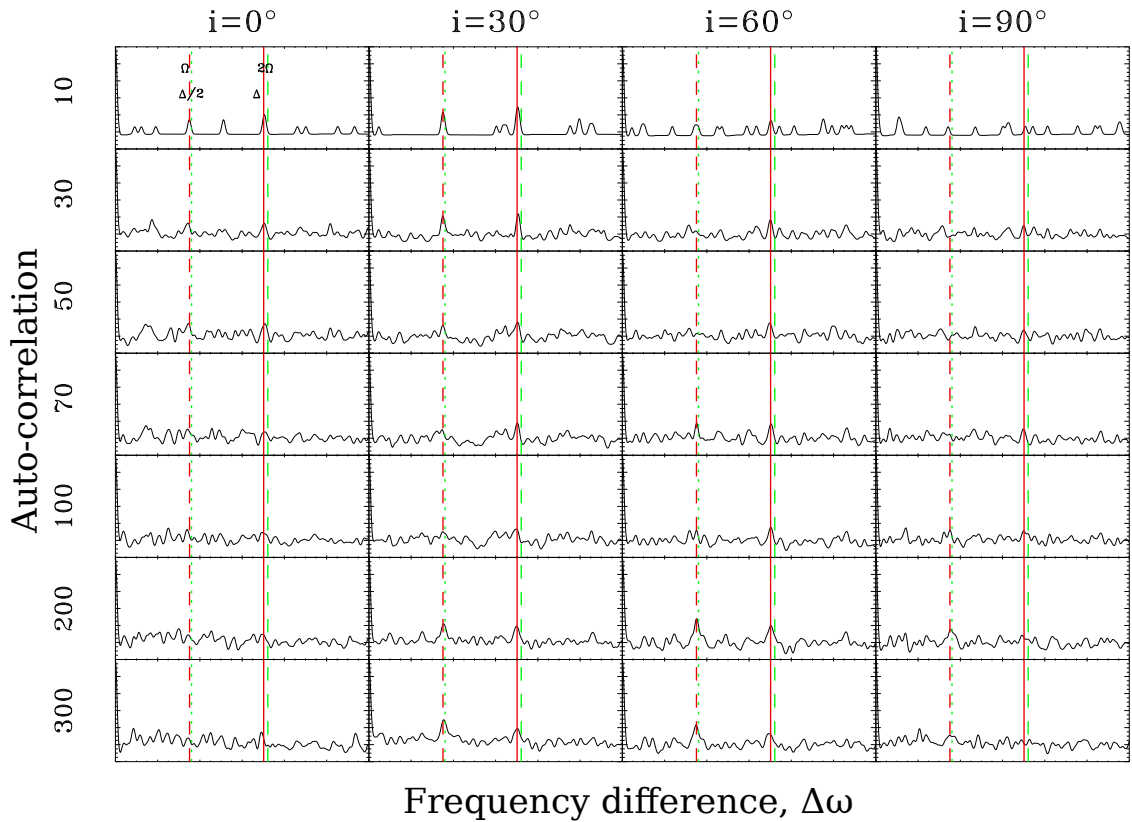}
\end{center}
\caption{(colour online) Same as Fig.~\ref{fig:auto5_7D} except that the model
is rotating at $0.7\,\OmegaC$, and the visibilities have been multiplied by
random numbers between $1$ and $100$ (which therefore affects which modes are
selected in the frequency spectra). \label{fig:auto7_7D_A100}}
\end{figure*}

Here, we multiplied the mode visibilities from the previous case by
random numbers, before selecting the highest-amplitude modes and calculating the
auto-correlation functions.  The random numbers are between 1 and 100 and are
uniformly distributed on a logarithmic scale.  Figure~\ref{fig:auto7_7D_A100}
shows the resultant auto-correlation functions for a model rotating at
$0.7\,\OmegaC$. As can be seen, the signature of regularities are much less
evident than previously. Nevertheless,  some peaks still remain, for instance
the peak around $\Delta$ for $i=30^{\circ}$ with $30$ modes and the peak around
$\Delta/2$, and to a lesser extent $\Delta$, for $i=60^{\circ}$ with $200$ or
$300$ modes. Similar signatures occurs around $0.3\OmegaC$ where $2\Omega$ is
close to $\Delta/2$. However, away from the coincidences between $2\Omega$ and
$\Delta$ or $\Delta/2$, the multiplication of the mode amplitude by such a
random factor  makes the characteristic frequency separations much more
difficult to extract.

\subsection{Other normalisations}

\begin{figure*}
\begin{center}
 \includegraphics[width=0.8\textwidth]{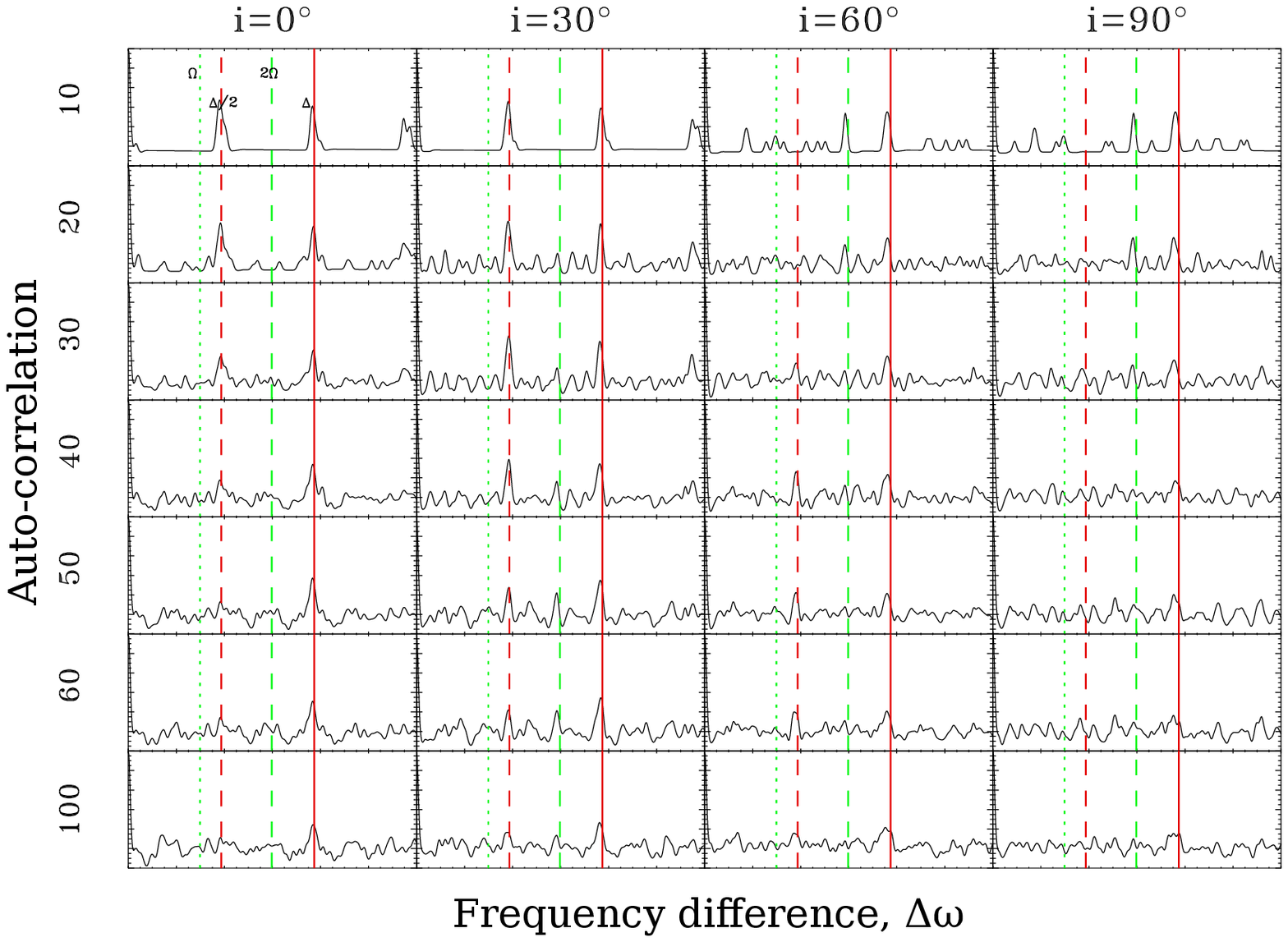}
\end{center}
\caption{(colour online) Same as Fig.~\ref{fig:auto5_7D} (\ie\ with a model
rotating at $0.5\OmegaC$) but where the modes are normalised by the kinetic
energy, multiplied by the appropriate power of $(\omega+m\Omega)$.
\label{fig:auto5_7D_bis}}
\end{figure*}

Figure~\ref{fig:auto5_7D_bis} shows how the auto-correlation functions
are modified when using the normalisation based on the kinetic energy. 
Qualitatively, this remains the same as Fig.~\ref{fig:auto5_7D}.  Some
of the peaks stand out better, notably the $2\Omega$ peaks for few modes. In
contrast, the normalisation based on the mean surface displacement (not shown)
gives poor results.  The explanation for this is quite simple: by normalising by
the mean surface displacement, gravity (or gravito-inertial) modes are no longer
penalised.  Furthermore, the $(\omega+m\Omega)^2$ factor in
Eq.~(\ref{eq:normalisation_surf}) ends up amplifying them.  Hence, the spectra
of selected modes based on this normalisation are dominated by gravity
modes, which do not follow the same pattern, thereby drowning out the $\Delta$,
$\Delta/2$, $\Omega$, and $2\Omega$ signatures in most cases.

\section{Fourier transform of the frequency spectrum}

\subsection{General description}

Recently, \citet{GarciaHernandez2009, GarciaHernandez2013} and
\citet{GarciaHernandez2015} analysed the Fourier transforms of the frequency
spectra of the $\delta$ Scuti stars HD 174936 \and{and HD 174966}, observed by
CoRoT.  They investigated what happens when the number of selected frequencies
varies from a few tens to a few hundreds.  In what follows, we apply the same
procedure but to our numerically calculated frequency spectra.  We
select modes according to their visibilities in CoRoT's photometric band, then
assign the same amplitude to the selected modes before calculating the Fourier
transform of the resultant spectrum.  In order to facilitate comparisons with
the auto-correlation functions, we apply this technique to the frequency spectra
spanning 7 radial orders, studied the previous section (\ie\
Figs.~\ref{fig:auto5_7D} and~\ref{fig:auto7_7D}).

\subsection{Normalisation of the maximal displacement}

\begin{figure*}
\begin{center}
 \includegraphics[width=0.8\textwidth]{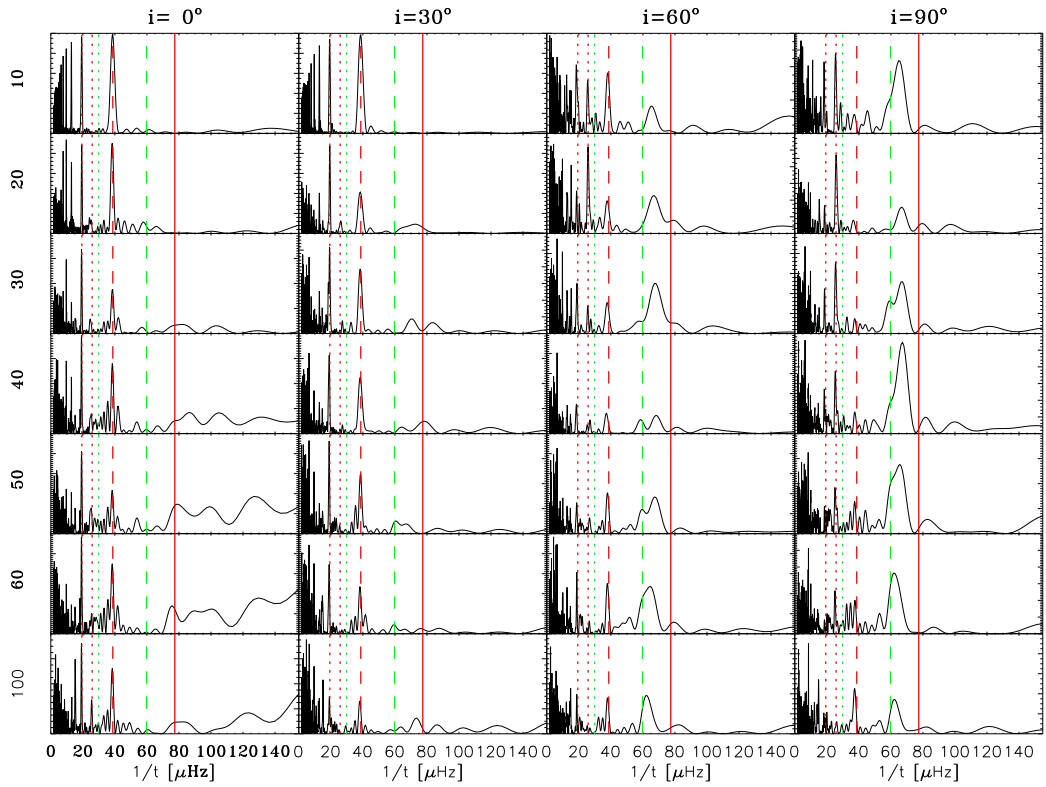}
\end{center}
\caption{(colour online) Fourier transform of frequency spectra in a
$2\,M_{\odot}$  model rotating at $0.5\,\OmegaC$. Each row corresponds to a
different number of selected modes, as indicated on the left, and each column to
a different inclination, as stated above.  The red, continuous and
discontinuous, vertical lines correspond to the large frequency separation,
$\Delta$, and various fractions of this value: $\Delta/4$, $\Delta/3$ and
$\Delta/2$.  The green, dotted and dashed, vertical lines correspond to $\Omega$
and $2\Omega$, respectively. \label{fig:TF5_7D}}
\end{figure*}

Figure~\ref{fig:TF5_7D} shows the squared modulus of the Fourier transform of
frequency spectra in the model rotating at $0.5\,\OmegaC$, for various numbers
of selected modes and for four different inclinations.  Taking
the Fourier transform of a function which depends on frequency yields another
function which depends on time $t$; it is plotted here as a function
of $1/t$ to facilitate the identification of regularities.  At low
inclinations, peaks appear at $\Delta/2$ with their forest of harmonics at
$\Delta/4$, $\Delta/6$, etc.  This can be expected because the frequency
spectrum is behaving like a Dirac comb with a $\Delta/2$ periodicity.   At
higher inclinations, a peak appears close to $\Delta$ (that is an expected
regularity), but is shifted; we also recover some harmonics (especially at
$\Delta/3$), but not all of them. This is probably an effect of rotation that
does not necessarily add peaks but acts as a modulation of the amplitude of the
Fourier transform. Indeed, we notice that rotation does not produce peaks at
$2\Omega$ or $\Omega$.  This is because, although there are recurrent frequency
separations of $2\Omega$ (or actually slightly smaller because of the Coriolis
force), such separations are formed by \textit{pairs} of frequencies rather than
by a Dirac comb. Nevertheless, when the frequency spectrum is dominated by two
similar subspectra, the second one being identical to the first one but shifted
by $2\Omega$ (this is what happens when $m=\pm1$ modes dominate), the Fourier
transform of the full spectrum will be the Fourier transform of the subspectrum
multiplied by $\cos^2(\Omega t)$. Such a modulation can make some peaks
disappear or slightly shift some broad peaks. In such a configuration, it is
impossible to unambiguously detect the correct large separation with the Fourier
transform only.

\begin{figure*}
\begin{center}
 \includegraphics[width=0.8\textwidth]{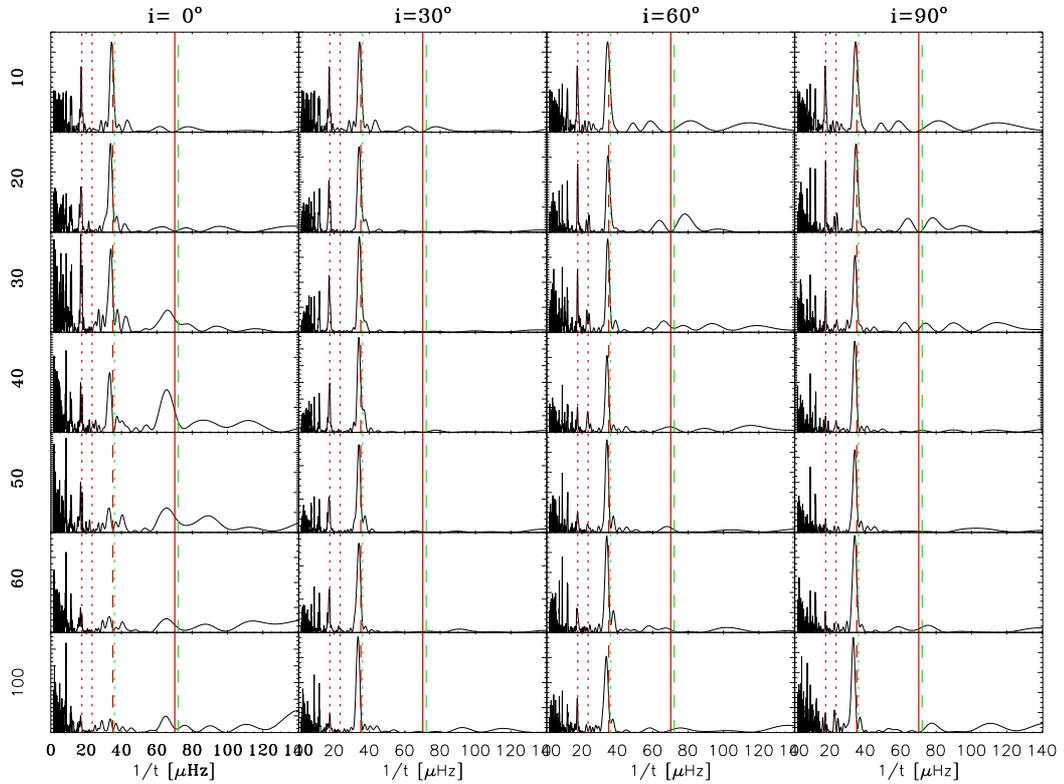}
\end{center}
\caption{(colour online) Same as Fig.~\ref{fig:TF5_7D} but for a model rotating
at $0.7\,\OmegaC$. \label{fig:TF7_7D}}
\end{figure*}

Figure~\ref{fig:TF7_7D} shows what happens with the model rotating at
$0.7\,\OmegaC$, where $\Delta$ nearly coincides with $2\Omega$.  In this case,
the frequency spectra take on a fairly simple form in which the frequencies
cluster around points separated by $\Delta/2 \simeq \Omega$, regardless of
inclination.  This leads to strong peaks at $\Delta/2$ and $\Delta/4$ in the
Fourier transforms, regardless of inclination. 

\subsection{Other normalisations}

\begin{figure*}
\begin{center}
\textbf{Inclusion of random factors ($\Omega=0.7\OmegaC$)} \\
 \includegraphics[width=0.8\textwidth]{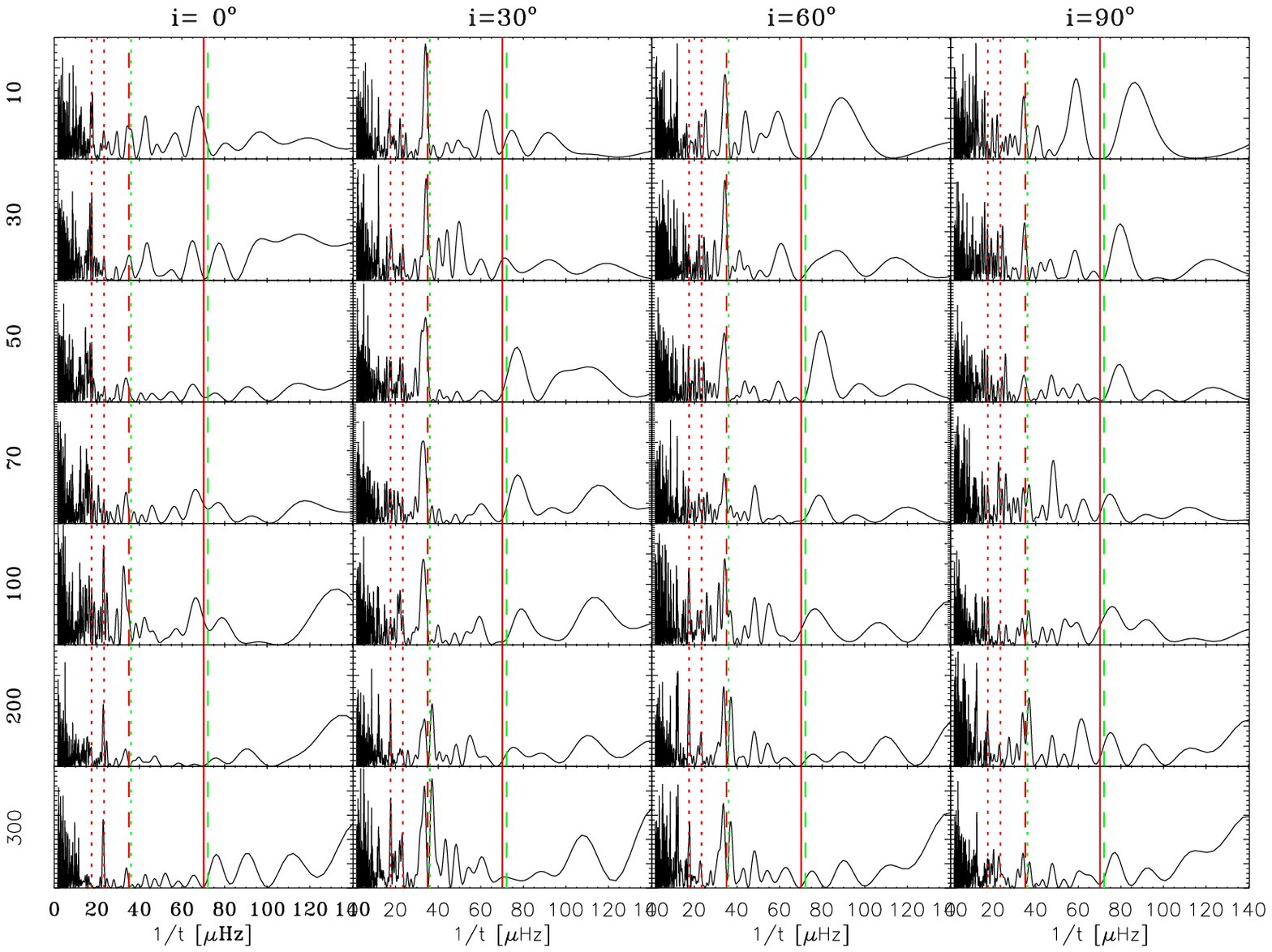} \\
\textbf{Mean surface displacement normalisation ($\Omega=0.5\OmegaC$)} \\
 \includegraphics[width=0.8\textwidth]{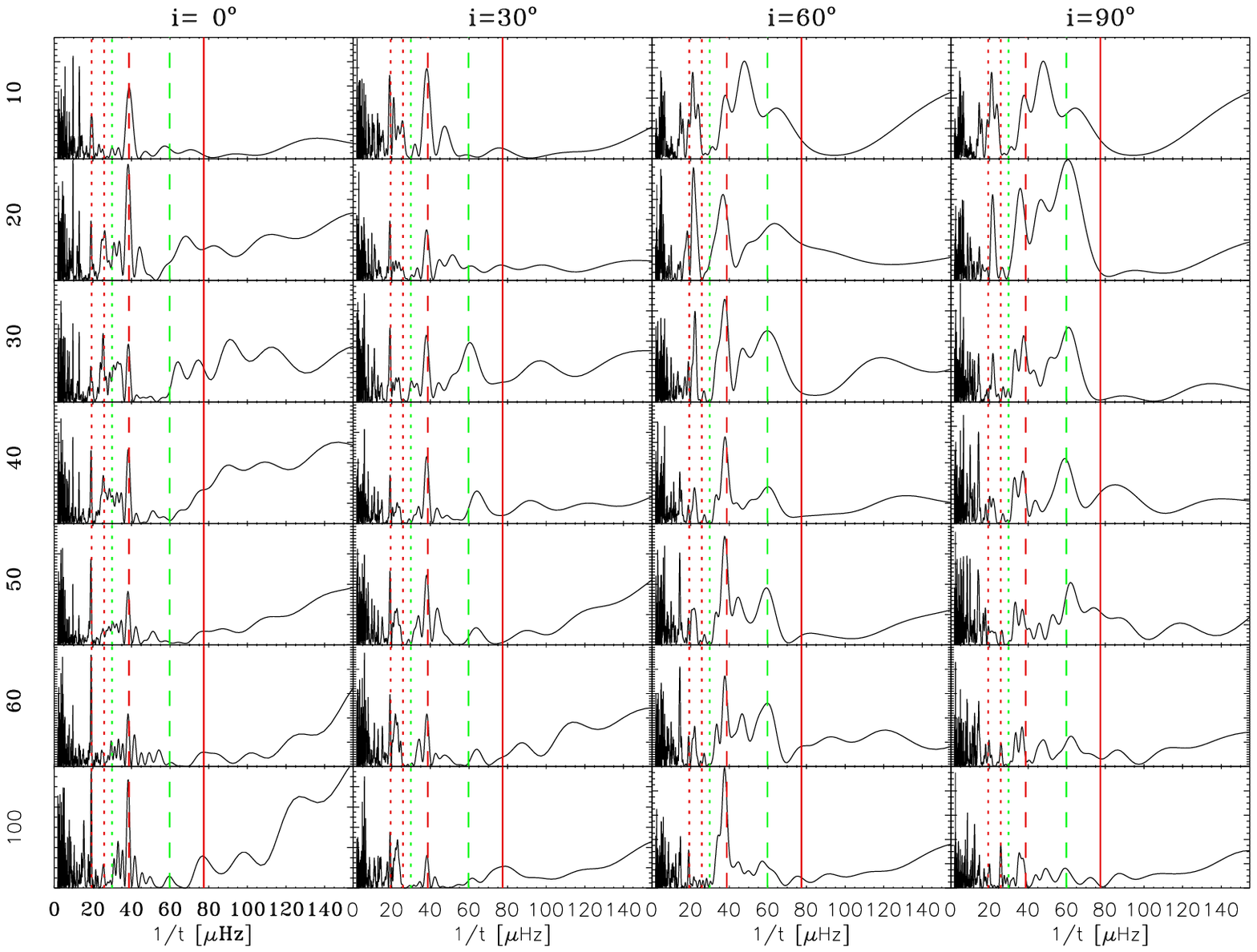}
\end{center}
\caption{(colour online) Same as Fig.~\ref{fig:TF5_7D} but where
the mode normalisation includes random factors (top panel)
or is based on the mean surface displacement (bottom panel).
\label{fig:TF5_7D_bis}}
\end{figure*}

Figure~\ref{fig:TF5_7D_bis} shows the effects of the alternative
normalisations described in Sect.~\ref{sect:normalisations} on the Fourier
transform of the spectra of the $\Omega = 0.5\OmegaC$ and $\Omega =
0.7\OmegaC$ models.  The Fourier transforms continue to detect $\Delta/2$ and
its many harmonics in many cases, even when using random factors or the
normalisation based on the mean surface displacement.  However, the
tests involving random factors benefit from the coincidence between $\Delta$ and
$2\Omega$.  In the absence of such a coincidence, the $\Delta/2$ signature is
far less visible, except in a few cases where it still shows up.
Overall, these tests confirm the robustness of the large and
semi-large frequency separations as detected by the Fourier transform
to different non-random normalisations, or when examining favourable
cases where $\Omega$ coincides with $\Delta/2$ or $\Delta$.

Overall, the Fourier transforms complement the auto-correlation functions quite
nicely.  Indeed, although both detect the separation $\Delta/2$, only the
auto-correlation functions are sensitive to $2\Omega$.  As explained above, such
frequency separations are produced by pairs of modes rather than by Dirac combs
thereby escaping detection by Fourier transforms, but not by the
auto-correlation functions.  Hence, this provides a simple way to distinguish
between the two and to get a better grasp of the regularities present in the
spectrum.

\section{Multi-colour mode identification}
\label{sect:multicolour}

\subsection{Method}

We now turn our attention to multi-colour mode identification.  As was
emphasised in the introduction, multi-colour mode signatures, such as the ratios
between mode amplitudes observed in different photometric bands, do not depend
on intrinsic mode amplitudes since these factor out.  In Paper~I, it was shown
that such amplitude ratios tend to be similar for island modes with the same
$(\l,\,|m|)$ values, even in the most rapidly rotating models.  This is
consistent with the fact that such modes have a similar surface structure (see
Section~\ref{sect:modes}). Hence, this raises the question as to whether we can
select modes with similar properties by picking a reference mode at random in an
oscillation spectrum and  searching for the $N$ other modes which produce the
most similar amplitude ratios. In what follows, we will use the following cost
function to evaluate the proximity of a mode's amplitude ratios to that of the
reference mode:
\begin{equation}
J = \sum_{i=1}^{d} \left( V_i - V_i^{\mathrm{ref}} \right)^2,
\label{eq:cost_mode_identification}
\end{equation}
where $i$ is an index on the photometric band, $d$ the number of photometric
bands, $V_i^{\mathrm{ref}}$ the \textit{visibilities} of the reference mode and
$V_i$ those of the mode being evaluated.   The visibilities have been
normalised using $\sum_{i=1}^{d} V_i^2 = \sum_{i=1}^{d}
\left(V_i^{\mathrm{ref}}\right)^2 = 1$ to avoid favouring a particular
photometric band.  Nonetheless, a small value for $J$ implies that the $V_i$'s
are close to the $V_i^{\mathrm{ref}}$'s and therefore that the amplitude ratios
are similar.

This approach is different than the one typically taken in other works
such as \citet{Daszynska_Daszkiewicz2002}.  Indeed, most authors compare
observed amplitude ratios directly with theoretical ones.  The approach
described here consists in comparing observed amplitude ratios between each
other. It thus bypasses limitations in the theoretical predictions. In the
following, we shall nevertheless use theoretical amplitude ratios to test its
validity.

\subsection{Adiabatic case}
\label{sect:multi-colour-adiabatic}

\begin{figure*}[htbp]
\begin{center}
 \includegraphics[width=0.9\textwidth]{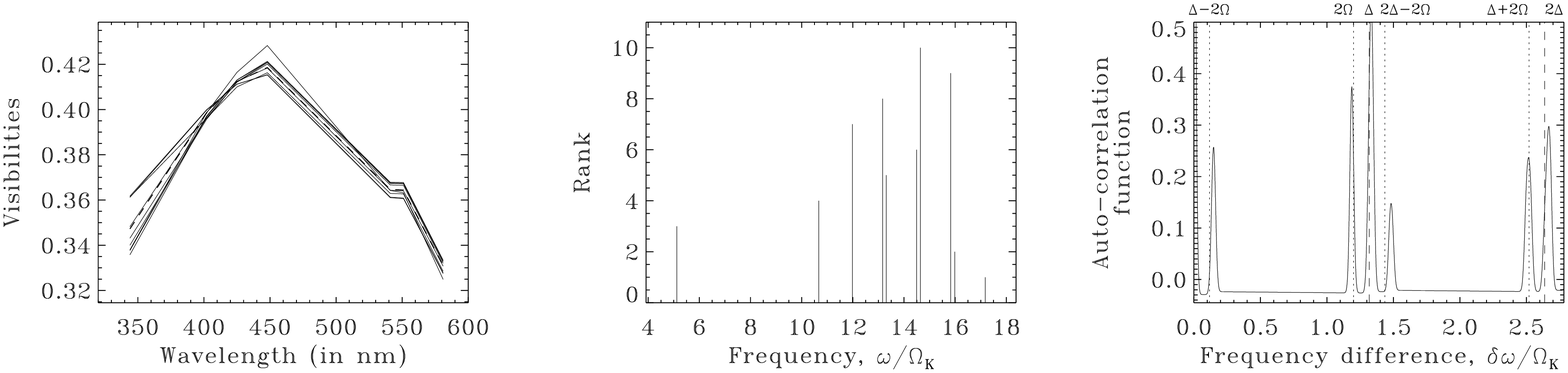}  \\
 \includegraphics[width=0.9\textwidth]{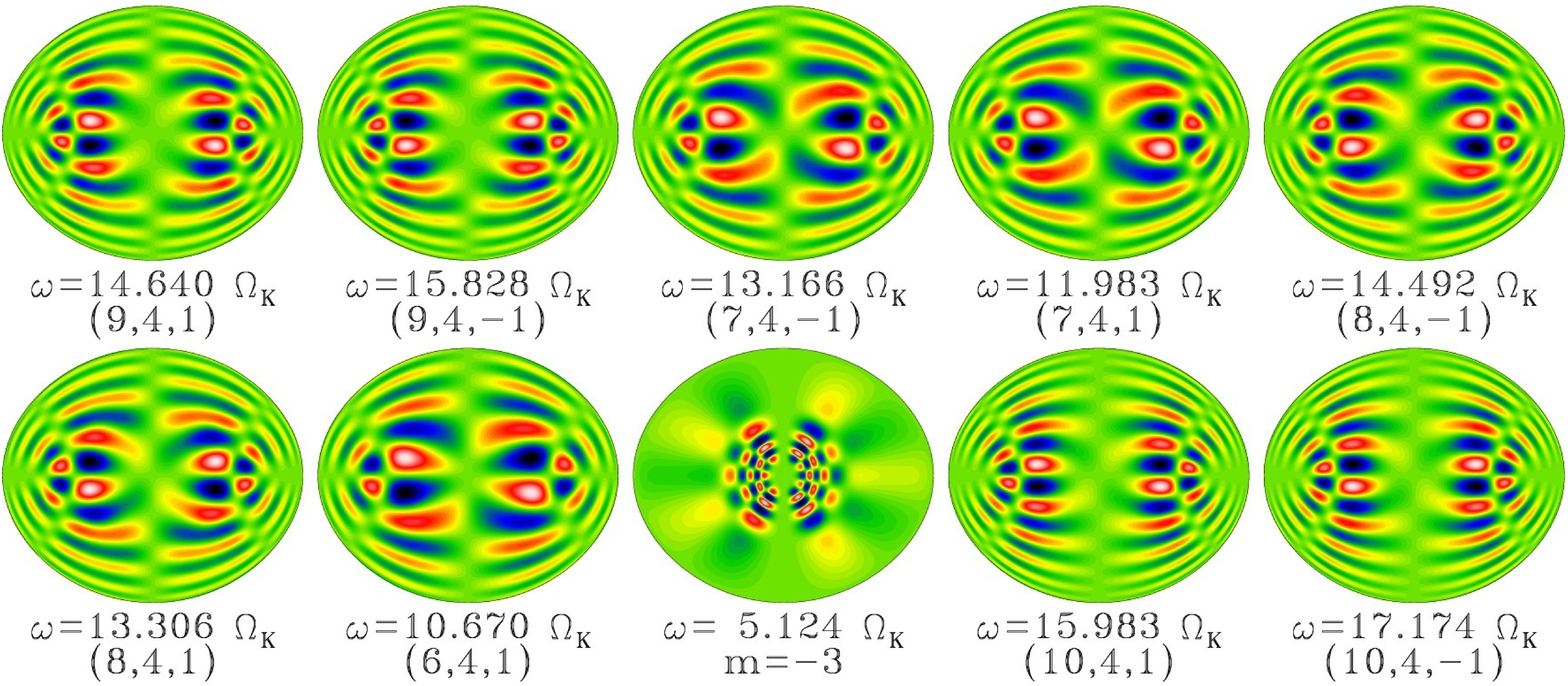}
\end{center}
\caption{(colour online) Amplitude ratios, frequency spectrum and ranks,
associated auto-correlation function, and meridional cross-sections of the
reference mode $(n,\,\l,\,m)=(9,4,1)$ and the $9$ other most similar modes, for
an inclination of $i=60^{\circ}$ -- see text for details.
\label{fig:match_case2}}
\end{figure*}

We start with the $(n,\,\l,\,m)=(9,4,1)$ mode in the model at
$0.6\,\OmegaC$ as the reference mode and search for the $9$ other modes with the
most similar amplitude ratios.  The mode visibilities are calculated using the
Geneva photometric system, which contains seven photometric bands. Before
normalising the visibilities, we filter out modes where the overall visibility
is more than 100 times smaller than that of the reference mode, thereby
excluding most gravity modes. Figure~\ref{fig:match_case2} shows the results
for $i=60^{\circ}$.  In the first row, the left panel shows the amplitude
ratios, the dashed line corresponding to the reference mode (we note that this
line is hardly visible because it is mostly covered up by the solid lines from
the other modes), the middle panel gives the frequencies, where a higher rank
indicates a higher resemblance with the reference mode, $10$ corresponding to
the reference mode, and the right panel displays the auto-correlation function
of the sub-spectrum of the selected frequencies. The second and third rows show
the meridional cross-sections of the selected modes.  Beneath each mode, the
values of $(n,\,\l,\,m)$ are indicated when the mode corresponds to an
$\tilde{\l} < 2$ island mode.  Otherwise, only the azimuthal order, $m$, is
provided.

The results shown in Fig.~\ref{fig:match_case2} are promising.  Apart
from the mode at  $5.124$ $\OmegaK$, all of the modes
belong to the $(\l,\,|m|)=(4,1)$ family of island modes. This then leads to
clear peaks in the auto-correlation function at $2\Omega$, $\Delta$, and
$2\Delta$, as well as combination peaks at $\Delta-2\Omega$, $2\Delta-2\Omega$,
and $\Delta+2\Omega$.  Another example of highly successful mode selection using
the same strategy can be found in Figure~4 of \citet{Reese2013b}.
However, not all cases work out so well.

To get a global picture of the mode selection method, we applied it to
all of the identified island modes above a given threshold frequency in the
pulsation spectrum of our 2 $M_{\odot}$, 0.6 $\OmegaC$ stellar model. Then, we
quantified the mode selection success rate by finding the average number of
island modes (excluding the reference mode), the average number of island modes
with the same $(\l,\,|m|)$ values as the reference mode, and the average number
of island modes with the same $(\tilde{\l},\,|m|)$ values, all of which
are subsequently divided by the total number of selected modes, \ie\
$N=9$, to get  a success rate comprised between $0$ and $1$.

Results are given in columns 2-4 of Table~\ref{tab:overall_success}, for
a threshold frequency $\omega_{\mathrm{threshold}}=8.0\,\OmegaC$. In this case,
the total number of island modes divided by the total number of modes is
$0.0115$. Thus, the much higher success rates obtained show that selecting modes
according to similar amplitude ratios considerably increases  the chances of
finding island modes, provided the reference mode is an island mode. When the
inclination is high, the method selects island modes of the same parity. In
contrast, in near pole-on configurations, it will select modes of both
equatorial symmetries. This property explains the difference in the $(\l,\,m)$
versus $(\tilde{\l},\,m)$ success rates. Repeating this test for different
values of the threshold frequency shows that higher success rates are achieved
for higher frequency modes, most probably because they are further into the
asymptotic regime.

\begin{table*}[htbp]
\begin{center}
\caption{Overall success rate in finding similar modes via the multi-colour mode
identification scheme. \label{tab:overall_success}}
\begin{tabular}{lccccccccc}
\hline
\hline
&
\multicolumn{3}{c}{\dotfill \textbf{Adiabatic (2 M$_{\odot}$)} \dotfill} &
\multicolumn{3}{c}{\dotfill \textbf{Adiabatic (1.8 M$_{\odot}$)} \dotfill} &
\multicolumn{3}{c}{\dotfill \textbf{Pseudo non-adiabatic (1.8 M$_{\odot}$)} \dotfill} \\
\textbf{Orientation} &
\textbf{Island} &
\textbf{Same} $(\l,\,|m|)$ &
\textbf{Same} $(\tilde{\l},\,|m|)$ &
\textbf{Island} &
\textbf{Same} $(\l,\,|m|)$ &
\textbf{Same} $(\tilde{\l},\,|m|)$ &
\textbf{Island} &
\textbf{Same} $(\l,\,|m|)$ &
\textbf{Same} $(\tilde{\l},\,|m|)$ \\
\hline
\textit{All}   & 0.564 & 0.359 & 0.416 & 0.554 & 0.401 & 0.452 & 0.469 & 0.303 & 0.349 \\
$i=0^{\circ}$  & 0.519 & 0.241 & 0.519 & 0.594 & 0.256 & 0.594 & 0.633 & 0.256 & 0.633 \\
$i=30^{\circ}$ & 0.474 & 0.268 & 0.375 & 0.483 & 0.343 & 0.430 & 0.441 & 0.303 & 0.357 \\
$i=60^{\circ}$ & 0.643 & 0.444 & 0.450 & 0.600 & 0.449 & 0.449 & 0.485 & 0.323 & 0.340 \\
$i=90^{\circ}$ & 0.597 & 0.402 & 0.402 & 0.592 & 0.462 & 0.462 & 0.450 & 0.275 & 0.275 \\
\hline
\end{tabular}
\end{center}
\end{table*}

Overall, the method appears to be a promising way of choosing
modes with similar surface distributions and hence quantum numbers. Regularities
of the sub-spectrum  of selected modes may then help to determine the azimuthal
order, thanks to the frequency separations $2|m|\Omega$, and constrain the
radial order.  Nonetheless, one may still wonder if such a strategy will
continue to work when non-adiabatic effects are taken into account.  In what
follows we examine this question by analysing mode visibilities in which
non-adiabatic effects are approximated.

\subsection{Pseudo non-adiabatic effects}

Non-adiabatic effects strongly modify the effective temperature variations and,
hence, multi-colour photometric signatures of pulsation modes.  One may then
wonder if such effects are able to mask the structural similarities between
modes with similar quantum numbers and thus alter the promising results
of the mode selection method obtained in the adiabatic case. In the absence of
full non-adiabatic pulsation calculations in rapidly rotating $\delta$
Scuti stars, we shall use non-adiabatic calculations in the non-rotating case
to derive approximate non-adiabatic visibilities in the rotating case.

\begin{figure}[htbp]
 \includegraphics[width=0.45\textwidth]{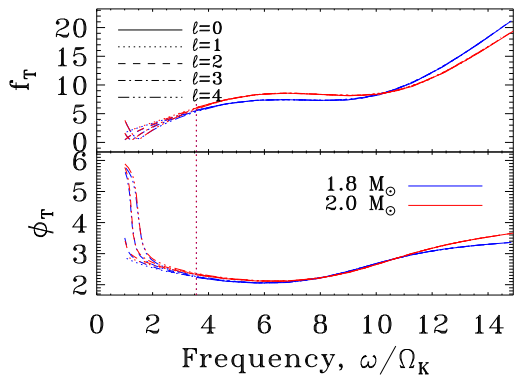}
\caption{(Colour online) The quantities $f_T$ and $\phi_T$ in two non-rotating
models with masses 1.8 $M_{\odot}$ and 2.0 $M_{\odot}$. The two models nearly
have the same temperature (8143.9 K and 8137.3 K, respectively), but different
$\log(g)$ values (4.1249 dex and 3.9593 dex, respectively).  The different line
styles correspond to different $\l$ values, and the overlapping vertical dotted
lines (which are hard to distinguish) indicate the position of the fundamental
modes of both models. \label{fig:fT}}
\end{figure}

\begin{figure}[htbp]
\begin{center}
 \includegraphics[width=0.43\textwidth]{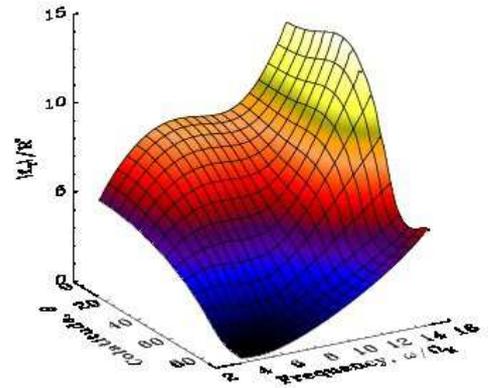}
\end{center}
\caption{(Colour online) The quantity $|f_T|/R^{\star}$ as a function of
co-latitude and frequency, for the 1.8 $M_{\odot}$ rotating model.
\label{fig:3D_fT}}
\end{figure}

In non-rotating stars, the relative effective temperature fluctuations are
typically proportional to the radial displacement:
\begin{equation}
\frac{\delta \Teff}{\Teff} = \frac{\xi_r}{R^{\star}} f_T \exp(i \phi_T)
\end{equation}
where $R^{\star}$ is the radius of the non-rotating model.  The quantity
$f_T\exp(i\phi_T)$ represents the non-adiabatic effects and
generally depends on the degree $\l$ and frequency $\omega/\OmegaK$ of the
mode, as well as on the effective temperature and surface gravity. However, as
illustrated in Fig.~\ref{fig:fT}, it actually depends little on the
harmonic degree for acoustic modes (\ie\ at sufficiently high frequencies).  We
also note that, when described as a function of $\omega/\OmegaK$,
$f_T\exp(i\phi_T)$ is only slightly affected by the effective gravity as can be
seen by comparing the two models in Fig.~\ref{fig:fT}, except at high
frequencies, where the effect is somewhat stronger. In a
rotating star, the effective temperature and the surface gravity vary from pole
to equator. Hence, to estimate non-adiabatic effects, we calculated, using the
MAD code \citep{Dupret2001}, $f_T\exp(i\psi_T)/R^{\star}$ for a set of
non-rotating models along four evolutionary sequences, which span the effective
temperature and surface gravity ranges of our rotating models, and for
a set of modes spanning the relevant frequency range.  We smoothed out the small
differences in degree and interpolated according to frequency, effective
temperature, and effective gravity, so as to have a value for
$f_T\exp(i\phi_T)/R^{\star}$ at each co-latitude for each frequency. The
function $|f_T|/R^{\star}$ is shown in Fig.~\ref{fig:3D_fT}.  This was
subsequently multiplied by $R(\theta)$, the radius of the rotating model which
depends on the co-latitude, $\theta$, and then applied at the corotating
pulsation frequency, $(\omega+m\Omega)/\Omega_K$ in order to estimate the
effective temperature variations arising in our pulsating rotating models.
A quick look at the function $|f_T|/R^{\star}$ shows that it cannot
simply be expressed as the product of a function which depends on $\theta$ alone
and another function which depends on frequency alone, \ie\ it is non-separable
in terms of these two variables.  Accordingly, this will distort the $\theta$
dependence of $\delta \Teff/\Teff$ as the frequency changes, thereby leading to
increased scatter in the photometric signatures of a given $(\l,\,|m|)$ family
of island modes.

\begin{figure*}[htbp]
\begin{center}
\textbf{Adiabatic case} \\
 \includegraphics[width=0.9\textwidth]{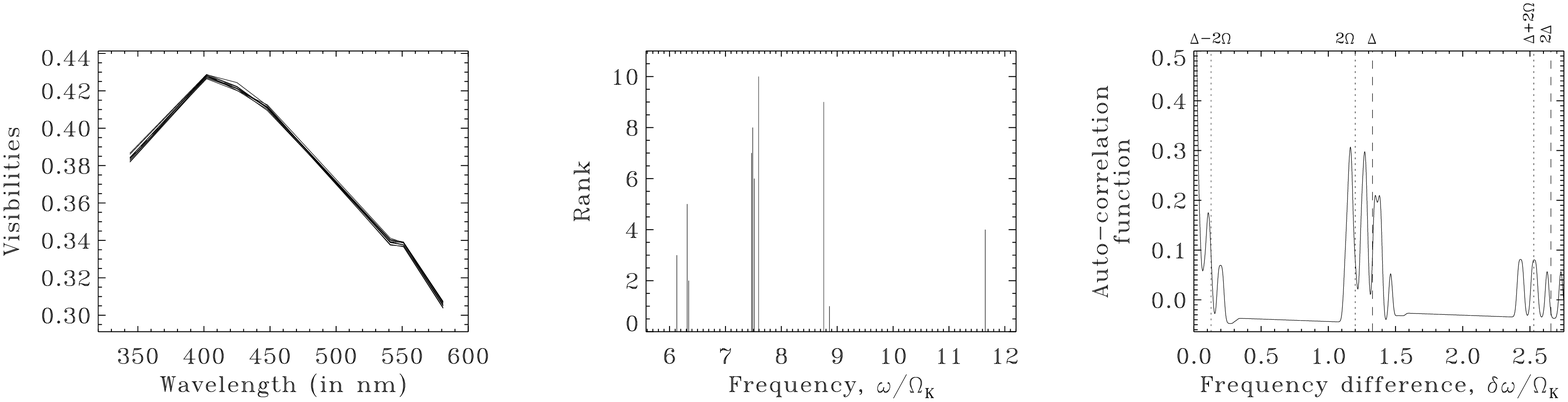}  \\
 \includegraphics[width=0.9\textwidth]{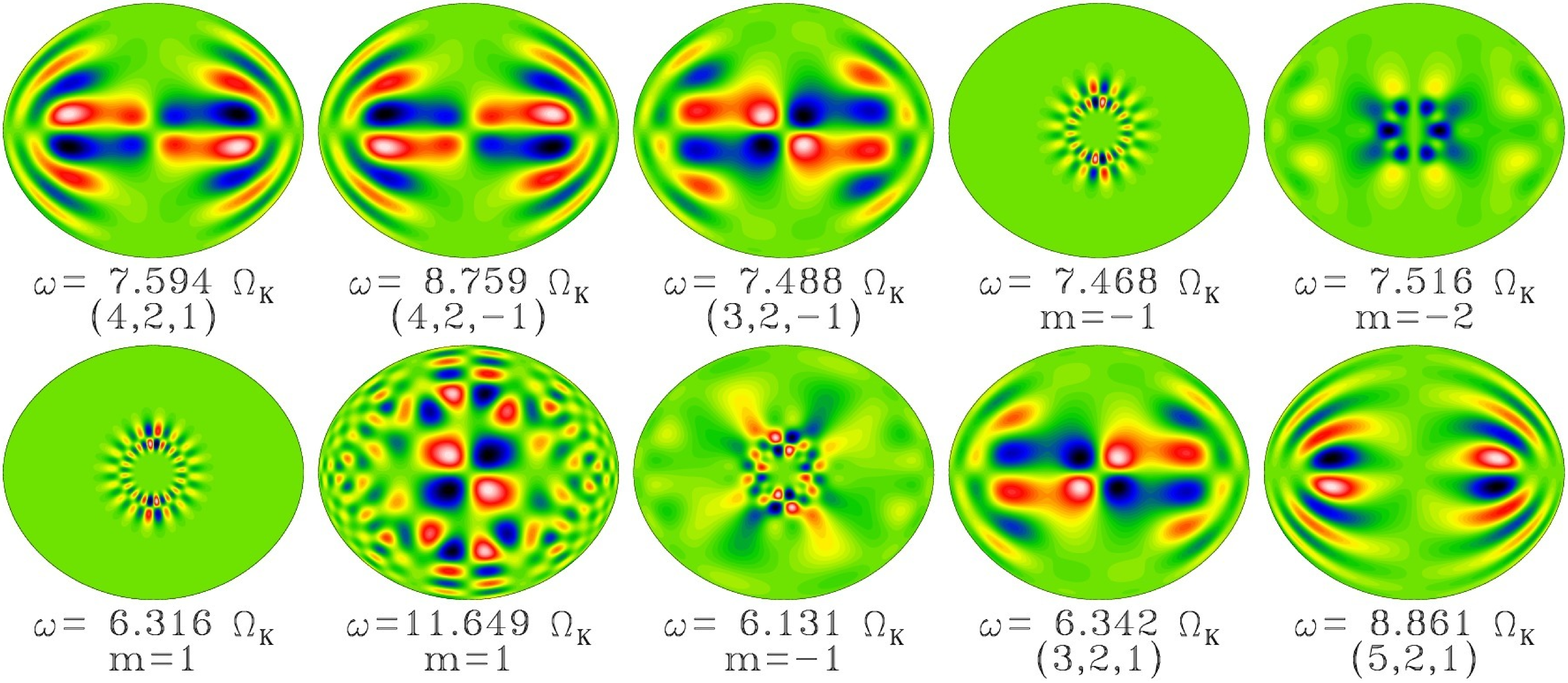} \\
\textbf{Pseudo non-adiabatic case} \\
 \includegraphics[width=0.9\textwidth]{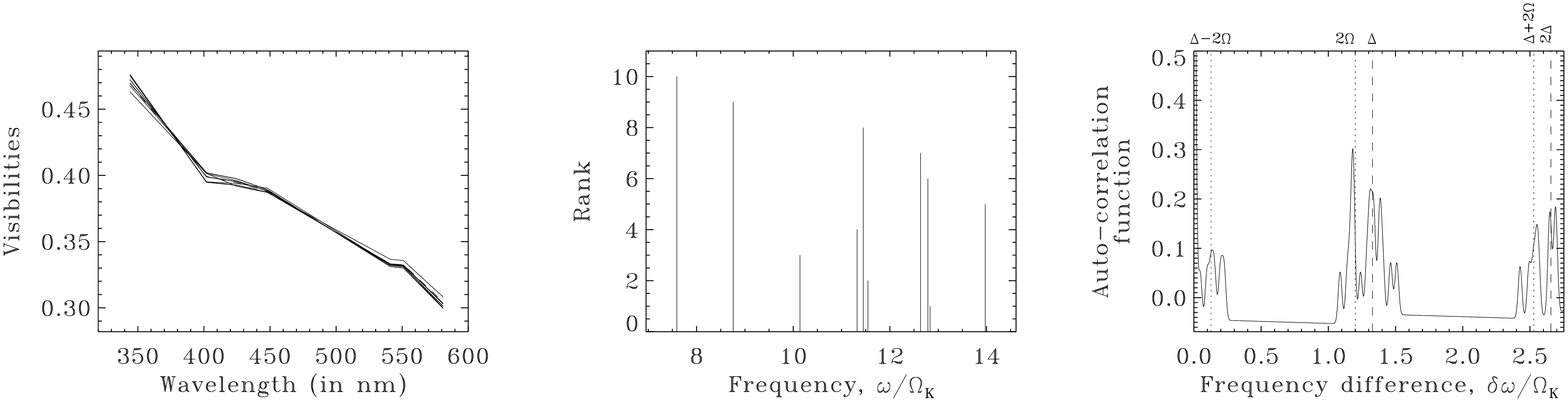}  \\
 \includegraphics[width=0.9\textwidth]{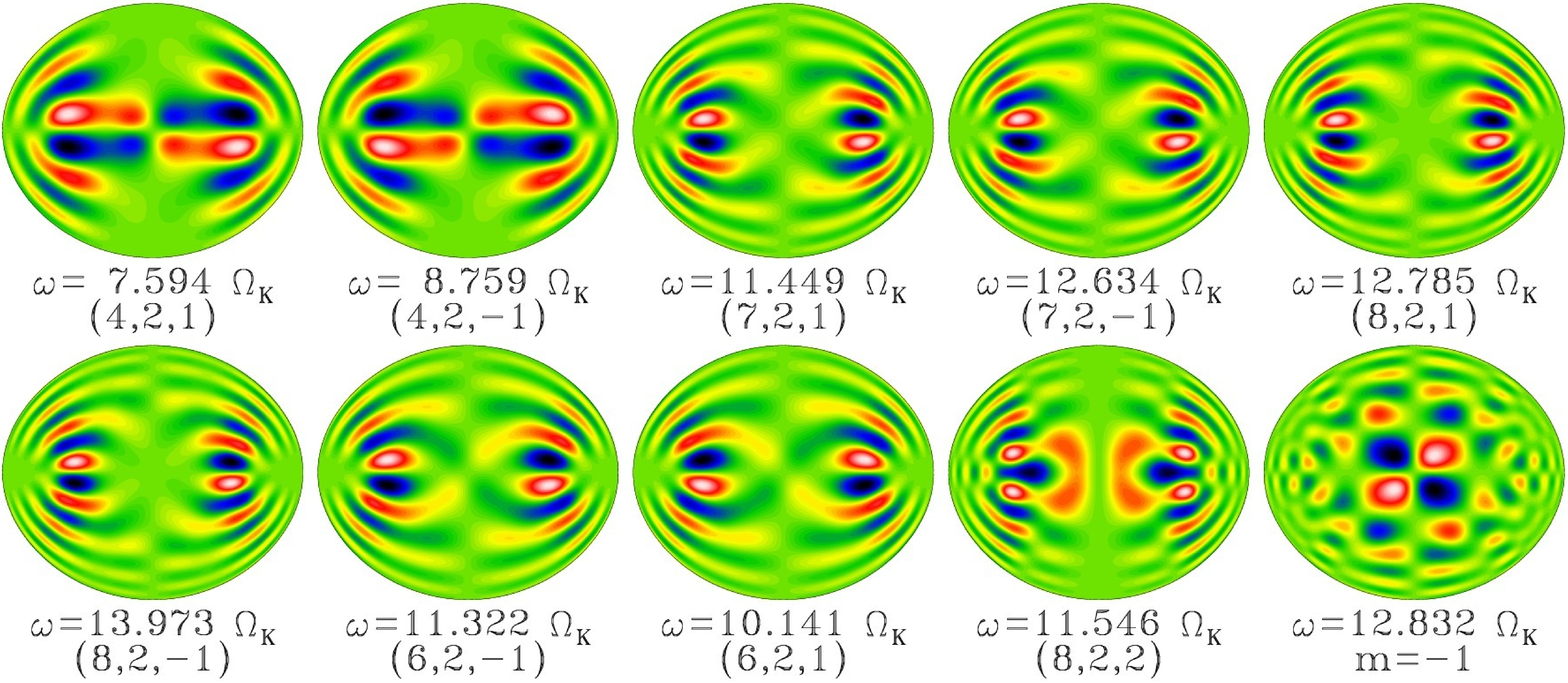}
\end{center}
\caption{(Colour online) Same as Fig.~\ref{fig:match_case2}
but for the reference mode $(n,\,\l,\,m)=(4,2,1)$ \textbf{(and for
both the adiabatic and pseudo non-adiabatic case)}. \label{fig:match_case3_fT}}
\end{figure*}

As in the previous section, the method to find similar modes is applied
to all of the island modes above a frequency threshold of $8.0\,\OmegaC$ and its
efficiency is quantified by  computing an overall success rate. In
Table~\ref{tab:overall_success}  the adiabatic results for a rotating model, $M
= 1.8\,M_{\odot}$, $\Omega = 0.6\,\OmegaC$, (see columns 5-7) are compared  with
the pseudo non-adiabatic ones (columns 8-10). In the present case, the number of
island modes divided by the total number of modes is $0.0330$. 
Table~\ref{tab:overall_success} shows  that pseudo non-adiabatic success rates
are worse than adiabatic ones but the method remains efficient in selecting
island modes, when island modes are used as reference modes.

There are a number of cases where pseudo non-adiabatic calculations lead
to similar or even better results than the adiabatic calculations, such as the
case illustrated in Fig.~\ref{fig:match_case3_fT}. In the adiabatic case, five
of the selected modes are not island modes, even though the reference mode is an
island mode.   In the pseudo non-adiabatic case, only one of the selected modes
is not an island mode, and there is also one island mode with a different value
of $|m|$.  The remaining modes correspond to pairs of $(\l,|m|) = (2,1)$ island
modes.  This gives rise to strong signatures at $2\Omega$, $\Delta$ and
$2\Delta$, as well as some combination peaks in the auto-correlation function.

Hence the technique appears to remain viable in the non-adiabatic case.
Furthermore, phase differences \citep[\eg][]{Balona1999b,Dupret2005} could also
be exploited and may lead to higher success rates by providing supplementary
constraints. This can, however, only be tested when full non-adiabatic pulsation
calculations are available in rapidly rotating $\delta$ Scuti stars, since it is
only in such conditions that phase differences can be obtained with a reasonable
accuracy.

\section{Conclusion}

In this paper, we investigated different ways of applying the visibility
calculations of \citet{Reese2013} to the interpretation of pulsation spectra in
rapidly rotating stars.  Given the lack of a comprehensive theory on
non-linear mode saturation in $\delta$ Scuti stars, we tested various ad-hoc
assumptions in order to determine intrinsic mode amplitudes.  As such, these
results must be taken with caution but represent a first step towards
interpreting pulsation spectra in such stars using realistic mode visibilities.

We first looked at the auto-correlation functions of frequency spectra in much
the same way as had been done in \citet{Lignieres2010}, but using the newer more
realistic visibilities, applied to spectra of acoustic modes calculated in a
self-consistent way (rather than having the non-axisymmetric modes estimated
from the axisymmetric ones).  Our results confirm those of \citet{Lignieres2010}
in the sense that it is possible to observe peaks corresponding to a mean value
of the large frequency separation, $\Delta$, half its value, $\Delta/2$, the
rotation rate, $\Omega$, and twice the rotation rate, $2\Omega$, when conditions
are favourable.  Such conditions are achieved when the frequencies span a large
enough range to reinforce recurrent spacings, and when not too many modes are
included.  The orientation of the star is important as it will favour either
$\Delta/2$ for low inclinations (\ie\  close to pole-on), or $\Delta$ and
$2\Omega$ for high inclinations.  The $\Omega$ spacing is visible due to the
usual multiplet structure at small rotation rates,  then it disappears and
becomes visible again at high rotation rates when a new of type multiplet
forms.  Of particular interest are the situations where $2\Omega$ coincides with
either $\Delta/2$ or $\Delta$, which occurs for models rotating at
$0.3\,\OmegaC$ or $0.7\,\OmegaC$, respectively.  Such situations lead to a very
strong signature in the auto-correlation function, due to a simplification of
the spectrum in which modes tend to cluster together.  Although this is ideal
for detecting those frequency separations, it also makes it more difficult to
disentangle the two.  Finally, we experimented with multiplying the intrinsic
amplitudes by random numbers between $1$ and $100$, uniformly spread out on a
logarithmic scale.  This is used as a poor substitute for the effects of
non-linear saturation and mode coupling on the amplitudes in such stars.  In
spite of that, a weak signature of the frequency separations remained in the
favourable case where $\Delta$ coincided with $2\Omega$.  This gives hope that
it may be possible to detect, at least in favourable cases, such characteristic
separations, and might explain the recent detection of recurrent
spacings in $\delta$ Scuti stars \citep{Mantegazza2012, Suarez2014,
GarciaHernandez2009, GarciaHernandez2013, GarciaHernandez2015}.

We also looked at Fourier transforms of the frequency spectra.  Although these
also detected $\Delta/2$ as a recurrent spacing, $2\Omega$ escaped detection. 
This is because the latter are not formed by a Dirac comb, but rather by
isolated pairs of frequencies.  This difference between the two approaches is
quite useful as it helps us distinguish between the two types of separations.
Hence, one can hope to interpret observed spectra by combining the two
approaches and correctly identifying $\Delta/2$ and $2\Omega$. Further tests
based on different mode normalisations confirmed the robustness of both the
auto-correlation functions and the Fourier transforms at detecting these
characteristic frequency separations, as long as gravito-inertial modes were
excluded from the analysis.

Finally, we turned our attention to multi-colour mode identification.  A key
advantage of multi-colour mode signatures, such as amplitude ratios, is that the
intrinsic amplitudes factor out, thereby leaving a signature which is only
sensitive to the geometric structure of the mode.  Previous investigations into
the matter had concluded that, due to the dependence of amplitude ratios on
inclination and azimuthal order, it would be very difficult to identify modes in
rapidly rotating stars from multi-colour photometry alone
\citep[\eg][]{Daszynska_Daszkiewicz2002, Townsend2003b}.  In contrast, we
present more promising results.  To achieve this, we apply a strategy different
from the non-rotating case, a strategy which involves choosing a reference mode
and searching for other modes with the most similar amplitude ratios.  By
repeating this procedure for different reference modes, we can group modes
together into families with similar amplitude ratios.  It turns out that such
families have similar azimuthal orders and degrees, and their frequencies follow
patterns which help to constrain the identification of these modes as well as
the rotation rate and the large frequency separation.  Furthermore, by comparing
modes between each other rather than with theoretical predictions, one bypasses
the current limitations with our theory such as the lack of full 2D
non-adiabatic pulsation calculations, or the difficulties in modelling the
interactions between such modes and convection.  Nonetheless, we investigated,
in an approximate way, non-adiabatic effects on multi-colour mode amplitudes. 
It was found that non-adiabatic effects do tend to increase the scatter between
the multi-colour photometric signatures of similar modes due to its distortion
of $\delta\Teff/\Teff$ as a function of frequency.  This makes the above mode
identification strategy more difficult to apply although some of the results
still remain promising.  Even if it turned out to be too penalising for mode
identification, it would still be possible to extract recurrent spacings such as
the rotation rate, and possibly the large frequency separation.

Of course, for multi-colour mode identification to work well, one needs high
quality multi-colour photometric observations of stars with numerous pulsation
modes, such as $\delta$ Scuti stars.  In this regard, the constellation of
nano-satellites, BRITE \citep{Kuschnig2009}, is a promising source of such data,
given that it observes in red and blue photometric bands.  The PLATO mission,
scheduled for launch in $2025$, will contain a platform of
$26$ telescopes, two of which will include broadband filters and the
remaining $24$ will operate in white light \citep{Catala2011,
Rauer2014}.


\begin{acknowledgements}

The authors thank the third referee for clear recommendations and
suggestions which have helped improve the manuscript.
DRR was financially supported through a postdoctoral fellowship from the
``Subside fédéral pour la recherche 2012'', University of Liège, and
was funded by the European Community's Seventh Framework Programme
(FP7/2007-2013) under grant agreement no. 312844 (SPACEINN), both of which are
gratefully acknowledged. This work was granted access to the HPC resources of
IDRIS under the allocation 2011-99992  made by GENCI (Grand Equipement National
de Calcul Intensif).

\end{acknowledgements}

\bibliographystyle{aa}
\bibliography{biblio}

\end{document}